\documentclass[reprint,aps,amsmath,amssymb,prfluids,onecolumn,11pt,floatfix]{revtex4-2}
\usepackage{graphicx}
\usepackage{natbib}
\usepackage{tabularx}
\usepackage{subcaption}
\usepackage{amsmath}
\usepackage{bbm}
\usepackage{tikz}
\usepackage[normalem]{ulem}
\usepackage{xcolor}
\usepackage{setspace} 

\colorlet{revision}{black}
\colorlet{houssem}{red}
\definecolor{gpltRed}   {HTML}{DC143C}
\definecolor{gpltBlue}  {HTML}{4169E1}
\definecolor{gpltGreen} {HTML}{6B8E23}
\definecolor{gpltGold}  {HTML}{DAA520}
\definecolor{gpltPurple}{HTML}{9932CC}

\definecolor{gpltAlpha01} {HTML}{AEE56C}
\definecolor{gpltAlpha03} {HTML}{679929}
\definecolor{gpltAlpha06} {HTML}{3D7F6A}
\definecolor{gpltAlpha12} {HTML}{0F4CBD}

\newcommand{\Rey}{\mathrm{Re}}
\newcommand{\Sto}{\mathrm{St}}
\newcommand{\Tau}{\mathrm{T}}

\DeclareRobustCommand\LineStyleSolid{\tikz[baseline]     \draw[]          (0,0.5ex) --++ (12pt,0);}
\DeclareRobustCommand\LineStyleSolidThick{\tikz[baseline]\draw[very thick](0,0.5ex) --++ (12pt,0);}
\DeclareRobustCommand\LineStyleDashed{\tikz[baseline]    \draw[dashed]    (0,0.5ex) --++ (12pt,0);}
\DeclareRobustCommand\LineStyleDashDot{\tikz[baseline]   \draw[dash dot]  (0,0.5ex) --++ (12pt,0);}

\DeclareRobustCommand\LineStyleSolidSquare{\tikz[baseline]{   \draw[] (0,0.5ex) --++ (12pt,0);\draw[black,fill=white,shift={(6pt,0.5ex)}] (-2.5pt,-2.5pt) rectangle (2.5pt,2.5pt);}}
\DeclareRobustCommand\LineStyleSolidCircle{\tikz[baseline]{   \draw[] (0,0.5ex) --++ (12pt,0);\draw[black,fill=white,shift={(6pt,0.5ex)}] (0,0) circle (2.5pt);}}
\DeclareRobustCommand\LineStyleSolidTriangleU{\tikz[baseline]{\draw[] (0,0.5ex) --++ (12pt,0);\draw[black,fill=white,shift={(6pt,0.5ex)}] (-2.5pt,-0.4*4.33pt) -- (2.5pt,-0.4*4.33pt) -- (0,0.6*4.33pt) -- cycle;}}
\DeclareRobustCommand\LineStyleSolidTriangleD{\tikz[baseline]{\draw[] (0,0.5ex) --++ (12pt,0);\draw[black,fill=white,shift={(6pt,0.5ex)}] (-2.5pt,0.4*4.33pt) -- (2.5pt,0.4*4.33pt) -- (0,-0.6*4.33pt) -- cycle;}}

\begin{document}

\title{Turbulence modulation in dense liquid-solid channel flow}

\author{Jonathan S. Van Doren}
\affiliation{
  School for Engineering of Matter, Transport and Energy, Arizona State University, Tempe, AZ 85281, USA
}
\author{M. Houssem Kasbaoui}
\email{houssem.kasbaoui@asu.edu}
\affiliation{
  School for Engineering of Matter, Transport and Energy, Arizona State University, Tempe, AZ 85281, USA
}
\date{\today}

\begin{abstract}
We investigate the mechanisms by which inertial solid particles modulate turbulence and alter the fluid mass transport in dense turbulent liquid-solid flows. To this end, we perform Euler-Lagrange simulations at friction Reynolds number 180, particle friction Stokes number 7.9, particle-to-fluid density ratio 8.9, and particle volume fraction ranging from 1\% up to 12 \%. We show that the mechanisms underpinning the flow modulation are two fold: (I) the increase of the suspension's apparent kinematic viscosity with increasing solid volume fraction and (II) turbulence modulation through the particle feedback force. For solid volume fraction below 3\%,  the increase of the suspension's apparent kinematic viscosity by the disperse particles accounts for most of the flow modification, namely, the reduction of turbulent fluctuations, reduction of the bulk fluid velocity, and increase of friction coefficient. In denser channels, the particle feedback force leads to greater reduction of bulk fluid velocity and increase of friction coefficient than can be accounted for solely based on the increased apparent kinematic viscosity. In these cases, particle stress significantly alters the stress balance, to a point where it exceeds the Reynolds stress at solid volume fraction 12\%.
\end{abstract}

\maketitle

\section{Introduction}

\textcolor{revision}{Slurry flows are commonly found in industrial and chemical settings, where solid particles are dispersed at high volume fractions inn a carrier liquid. Such flows are complicated due to turbulence modulation by suspended additives and possibly settling.} The presence of additives within the fluid can alter the underlying fluid flow such as the skin friction, turbulent statistics and the balance of stresses \citep{tomsObservationsFlowLinear1949, lumleyDragReductionAdditives1969, gyrDragReductionTurbulent2013,  kasbaouiTurbulenceModulationSettling2019}.  Analyzing the fluid mass transport and turbulence modulation requires special care in slurries because the particle volume fraction is not negligible, and particle collisions are significant. \textcolor{revision}{In the present study, we investigate these effects and show the interplay between particles and carrier fluid on the global mass and momentum transport in the case where particle settling is negligible.}

\textcolor{revision}{Additives such as spherical particles can lead to turbulence modulations, but the nature of the modulation in wall-bounded flows is not yet well understood.} Particles that are unable to follow streamlines due to their inertia exert micro stresses upon the carrier fluid. In the case of gas-solid flows, \citet{kasbaouiRapidDistortionTwoway2019} showed that dilute particle concentrations of the order $10^{-4}$ can modulate anisotropic stresses and enhance turbulent kinetic energy dissipation in Homogeneously sheared turbulence (HST).  \citet{kasbaouiRapidDistortionTwoway2019} and \citet{kasbaouiTurbulenceModulationSettling2019} studied particle modulation in such flows, they found that for volume fractions $\overline{\alpha}_{p} = 10^{-6} - 10^{-3}$ and mass loading $M=O(1)$, particles may augment or attenuate the turbulent kinetic energy.  Augmentation or attenuation is dependent upon the particles inertia. The mechanism of turbulence modulation in HST was described by \citet{ahmedDirectNumericalSimulation2001} who showed that the injection of energy from particles causes a reverse cascade of scales. \textcolor{revision}{\citet{nicolaiSpatialDistributionSmall2013} experimentally studied a water-glass homogenous shear flow and observed strong particle clustering at small scales. \citet{gualtieriAnisotropicClusteringInertial2009} performed simulations of particle-laden HST and observed anisotropic particle clustering. \citet{battistaApplicationExactRegularized2018} studied HST using the Exact Regularized Point Particle method, and found that particles with Kolmogorov Stokes number order one suppressed turbulent kinetic energy, while heavier particles had a negligible effect upon turbulent kinetic energy. \citet{buchtaSoundTurbulenceModulation2019} showed a reduction in velocity fluctions in Eulerian-Lagrange simulations of high speed shear flows with increasing particle mass loading.} In wall-bounded turbulent flows, inertial particles may cluster in the near wall region, leading to significant modulation to shear stress and the stress balance. \citet{sardinaWallAccumulationSpatial2012,nilsenVoronoiAnalysisPreferential2013,yuanThreedimensionalVoronoiAnalysis2018} performed simulations of dilute particle-laden turbulent channel flows which showed that the particle concentration in the viscous region may exceed the mean concentration by one or two orders of magnitude. \citet{liNumericalSimulationParticleladen2001} presented evidence of decreased skin friction in simulations using a point particle method for particles with friction Stokes number $St^{+} = \tau_p u_{\tau}^2/\nu = 192 $ dispersed in a vertical channel at friction Reynolds number $\Rey_\tau=u_{\tau}h/\nu=125$, where $\tau_{p}$ is the particle response time, $u_{\tau}$ is the friction velocity, $h$ is the channel half height, and $\nu$ is the kinematic viscosity. In this configuration they found an increase in fluid mass flow rate of $5 \%$ for mass flow rates $M = 0.2$, which is equivalent to a reduction in skin friction drag. \citet{zhaoTurbulenceModulationDrag2010} showed that inertial particles with $St^{+} = 30$ at mass loading $M = 0.32$ increase the fluid mass flow rate by $15\%$ in a channel at $\Rey_{\tau} = 180$, although these results may not be representative of the statistically stationary state. A follow up study by \citet{zhouNonmonotonicEffectMass2020a} in the same configuration showed much lower drag reduction, at only $~ 0.2\%$ for $M = 0.4$ and $2.8\%$ for $M = 0.75$. \textcolor{revision}{\citet{costaInterfaceresolvedSimulationsSmall2020} performed particle resolved direct numerical simulations (PR-DNS) of a dilute particle-laden turbulent channel flow at $\Rey_{\tau} = 180$, and compared with point particle simulations. They observed that inertial particles with $St^{+} = 50$ with $M = 0.03367$ led to increased skin-friction drag of $10\%$. \citet{costaNearwallTurbulenceModulation2021} performed PR-DNS simulations of a semi-dilute particle-laden turbulent channel flow at $\Rey_{\tau} = 180$. They observed that inertial particles with $St^{+} = 50$ with $M = 0.34$ led to increased skin-friction drag of $16\%$. Notably, the high computational costs of PR-DNS required \citet{costaInterfaceresolvedSimulationsSmall2020,costaNearwallTurbulenceModulation2021}} to use a domain which may have been too small to capture particle structures and their interaction with near-wall coherent structures. \citet{dritselisDirectNumericalSimulation2016} showed significant modification of the Reynolds stress tensor in the presence of particles even at particle volume fractions as low as $10^{-5}$. \textcolor{revision}{\citet{gualtieriEffectStokesNumber2023} performed point-particle simulations of turbulent channels at $\Rey_{\tau} = 185$ and mass loading $M = 0.4$, while varying the density ratio and Stokes number. They found increased drag at all density ratios, with a reduction in the effect at the highest density ratio. When Stokes number is lowest they found the highest drag increase. \citet{capecelatroTransitionTurbulenceRegimes2018a} considered particle-laden vertical channel flow at $\Rey_{\tau} = 300$, and varied the Stokes number and mass loading. At the highest mass loading, $M = 20$, they observed relaminarization of the flow. \citet{gaoDirectNumericalSimulation2023} simulated a particle-laden open channel flow at $\Rey_{\tau} = 5186$ using the Eulerian-Lagrange approach with Stokes numbers 448 and 6. They observed reductions in the rms velocity fluctuations, Reynolds stress, and TKE. \citet{rohillaEffectChannelDimensions2023} performed large eddy simulations (LES) partice-laden vertical channel flows while varying the bulk Reynolds number, particle volume fraction, and channel half height. They found fluid fluctuations decreased with increasing particle volume fraction, with the flow eventually relaminarizing. Additionally, they reported increased turbulence attenuation with increasing channel half height. \citet{capecelatroMassLoadingEffects2015} used the Eulerian-lagrange approach to investigate a particle-laden turbulent channel flow at $\Rey_{\tau} = 630$. They observed increased skin friction with increasing particle volume fraction. \citet{wangInertialParticleVelocity2019a} compared simulations using two point particle methods to experimental date for a particle-laden turbulent channel flow. They showed that with increasing particle volume fraction the point-particle methods considered show significant discrepancies with experimental data.} Recently, \citet{daveMechanismsDragReduction2023} used the point particle method to simulate a semi-dilute turbulent channel at $\Rey_{\tau} = 180$ and found that the friction Stokes number controls whether the skin-friction drag will increase or decrease, while the mass loading determines the magnitude of the modulation. They further showed that the mechanism for the turbulent modulation is related to the stress balance, which is dependent upon the viscous stress, Reynolds shear stress, and the particle stress. The introduction of a particle stress can cause the flow to approach the laminar mass flow rate, which is greater than the turbulent mass flow rate.
 
The previously investigated gas-solid flows have high particle-fluid density ratios, leading to high mass loading even in the semi-dilute regime. Liquid-solid slurries, however, are characterized by high volume fractions and low particle-fluid density ratios. The high particle volume fractions leads to stronger coupling between the fluid and particle phases. Specifically, liquid-solid slurries may have significant modulating effects associated with volume exclusion and increased apparent viscosity. These effects can lead to significant changes in the effective suspension Reynolds number. In addition, the interaction between suspended particles and turbulent flow structures may lead to strong turbulence modulation. \citet{matasTransitionTurbulenceParticulate2003,loiselEffectNeutrallyBuoyant2013,yuNumericalStudiesEffects2013} showed a decrease in the critical Reynolds number for transition in wall bounded flows in the semidilute regime. \citet{shaoFullyResolvedNumerical2012a}  considered both neutrally buoyant and dense particles in the turbulent regime, up to volume fractions of $7\%$. They found a decrease in the streamwise fluid velocity fluctuations caused by the attenuation of streamwise vortices. Additionally, the dense particles sedimented and formed a rough boundary at the lower wall, with particles free to resuspend. \citet{vowinckelFluidParticleInteraction2014}  varied the Shields number by way of particle density and found different regimes. \citet{picanoTurbulentChannelFlow2015a}  considered neutrally buoyant particles in the fully turbulent regime using particle resolved direct numerical simulations up to $20\%$ volume fraction. They observed interaction between the finite-size particles and turbulent motion leading to an alteration of the near-wall turbulence regeneration process. In dense suspensions, \citet{picanoTurbulentChannelFlow2015a} found that the law of the wall is modified, the mean streamwise velocity profile is greatly decreased, turbulent fluctuations are suppressed, and the low-speed streaks are altered compared to a particle-free channel. Additionally, they observed an alteration in the friction Reynolds number caused by both the altered effective viscosity of the suspension and an increase in the friction velocity, indicating an additional dissipation mechanism.

As slurry channels are characterized by high volume fractions, high-fidelity simulations must be able to capture complex fluid-particle, particle-fluid, and particle-particle interactions, which is known as four-way coupling. The difficulty in measuring flow properties in a channel slurry experimentally makes numerical simulations an attractive alternative. Euler-Lagrange (EL) formulations provide a high-fidelity framework for simulating dilute, semi-dilute, and dense particle-laden flows. In EL methods particles are tracked in a Lagrangian frame while equations for the fluid phase are solved on an Eulerian grid. \citet{capecelatroEulerLagrangeStrategy2013} used EL formulations previously to simulate dense fluidized beds and \citet{arollaTransportModelingSedimenting2015} used EL methods to reproduce bedform regimes in slurries in pipe flows. This makes the EL method suitable for the high volume fraction collisional slurry channel.

In this paper, we investigate the effect of increasing solid volume fraction on mass transport and stress balance in a turbulent channel flow using EL simulations. Section \ref{sec:Math_Model} provides a description of the mathematical model used in the present simulations. In section \ref{sec:Num_experiments}, we describe the  configuration and discuss the non-dimensional numbers that control the flow. To establish a baseline for comparison, we review the flow statistics and stress balance in a particle-free fully developed turbulent channel in section \ref{sec:Stat_State}. In section \ref{sec:Volume_Frac}, we discuss the results from the particle-laden channel flow simulations. In comparison with a particle-free channel, we show that increasing the solid volume fraction leads to decreasing streamwise fluid velocity, reducing velocity fluctuations, and increasing particle stresses. Using a novel analysis of the stress balance and additional simulations of particle-free channels with matching apparent viscosity, we analyze the relative effects of increased apparent viscosity and particle feedback force on the reduction of the bulk velocity in the carrier fluid. Finally, we give concluding remarks in section \ref{sec:conclus}.

\section{Mathematical model and computational parameters} 
\subsection{Mathematical model}
\label{sec:Math_Model}
We use the volume-filtering approach of \citet{andersonFluidMechanicalDescription1967} and Euler-Lagrange methodology of \citet{capecelatroEulerLagrangeStrategy2013} to describe the dynamics of the channel slurry. The carrier phase is an incompressible fluid with density $\rho_f$ and dynamic viscosity $\mu_f$. The volume-filtered equations governing mass and momentum conservation of the fluid phase read
\begin{eqnarray}
    \frac{\partial}{\partial t} (\alpha_{f}  \rho_f)+\nabla \cdot \boldsymbol(\alpha_{f} \rho_f \boldsymbol{u}_{f}) &=& 0\label{eq:continuity},\\
    \frac{\partial}{\partial t}(\alpha_f \rho_f \boldsymbol{u}_{f})+\nabla \cdot (\alpha_f {\rho_f} \boldsymbol{u}_{f} \boldsymbol{u}_{f}) &=& \nabla \cdot \left( \boldsymbol{\tau}+\boldsymbol{R}_{\mu} \right) - \boldsymbol{F}_{p} +\alpha_{f} A\boldsymbol{e}_x\label{eq:momentum},
\end{eqnarray}
where $\alpha_f$ is the fluid volume fraction, $\boldsymbol{u}_{f}$ is the volume-filtered fluid velocity, ${\boldsymbol{\tau}}=-p\boldsymbol{I}+\mu [\nabla \boldsymbol{u}_{f}+\nabla \boldsymbol{u}_{f}^{T}-\frac{2}{3}(\nabla \cdot \boldsymbol{u}_{f})\boldsymbol{I}]$ is the resolved fluid stress tensor, $\boldsymbol{R}_{\mu}$ is the so-called residual viscous stress tensor, $\boldsymbol{F}_{p}$ is the momentum exchange between the particles and the fluid \textcolor{revision}{(see equation (\ref{eq:F_p}))} \citep{capecelatroEulerLagrangeStrategy2013}. The last term on the right hand side of equation (\ref{eq:momentum}) represents a forcing akin to a pressure-gradient that drives the flow in the streamwise direction $x$. \textcolor{revision}{Note that the forcing magnitude $A$, which is constant here, relates to the wall shear stress following $\tau_w = \overline{\alpha}_f Ah$} (see appendix \ref{sec:appendix_a}), where $\overline{\alpha}_f$ is the channel-averaged fluid volume fraction. 

The residual stress tensor $\boldsymbol{R}_{\mu}$ arises from filtering the point-wise stress tensor. Since it includes sub-filter scale terms, $\boldsymbol{R}_{\mu}$ requires closure. It is believed that this term is responsible for the apparent enhanced viscosity observed in viscous fluids containing suspended solid particles \citep{einsteinNeueBestimmungMolekueldimensionen1905,thomasTransportCharacteristicsSuspension1965,gibilaroApparentViscosityFluidized2007,capecelatroEulerLagrangeStrategy2013}. Based on this rationale, \citet{capecelatroEulerLagrangeStrategy2013} proposed a closure of the form 
\begin{equation}
	\boldsymbol{R}_{\mu} = (\mu^\star_{f}-\mu_f)[\nabla \boldsymbol{u}_{f}+\nabla \boldsymbol{u}_{f}^{T}-\frac{2}{3}(\nabla \cdot \boldsymbol{u}_{f})\boldsymbol{I}],
\end{equation}
where $\mu_f^\star$ is the suspension's apparent viscosity.  \citet{einsteinNeueBestimmungMolekueldimensionen1905} notably showed that $\mu_f^\star/\mu_f=1+2.5\alpha_p$ for uniform dilute suspensions, where the particle volume fraction $\alpha_p<0.01$. Later \citet{thomasTransportCharacteristicsSuspension1965} and \citet{gibilaroApparentViscosityFluidized2007} derived semi-empirical relations by fitting experimental data sets. These relations extend well above the limit of validity of Einstein's apparent viscosity, up to $\alpha_p\sim 0.4$. In the present study, we use the apparent viscosity model of \citet{gibilaroApparentViscosityFluidized2007} due to its compact form, which reads $\mu_f^\star/\mu_f=(1-\alpha_p)^{-2.8}$. This relationship reveals the first anticipated effect: compared to a particle-free turbulent flow, dispersing particles at the concentrations encountered in slurries would suppress turbulence as the suspension becomes increasingly more viscous with rising $\alpha_p$. 

{\color{revision}
The particles are described in the Lagrangian frame. Following \citet{maxeyEquationMotionSmall1983a}, the equations of motion of a particle ``$i$'' located at $\boldsymbol{x}^i_p$, moving at velocity $\boldsymbol{v}^i_p$ and angular velocity $\boldsymbol{\omega}^i_p$ are given by 
\begin{eqnarray}
    \frac{d \boldsymbol{x}^i_p}{d t}(t) &=& \boldsymbol{u}^i_p(t) \label{eq:lpt_1}\\
    m_p \frac{d \boldsymbol{u}^i_{p}}{d t}(t) &=& \boldsymbol{f}_{p}^{h,i}+ \boldsymbol{f}_{p}^{c,i} \label{eq:lpt_2}\\
    I_p \frac{d \boldsymbol{\omega}^i_{p}}{d t}(t) &=& \boldsymbol{T}_{p}^{c,i} \label{eq:lpt_2}\\
\end{eqnarray}
where $m_p=\rho_p\pi d_p^3/6$ is the mass of a particle with density $\rho_p$ and diameter $d_p$, $I_p$ is its moment of inertia , $\boldsymbol{f}_p^{c,i}$ and $\boldsymbol{T}_p^{c,i}$ represents the collisional force and torque exerted on the particle due to particle-particle and particle-wall collisions.

The hydrodynamic force on the particles is modeled using
\begin{equation}
 \boldsymbol{f}^{h,i}_{p}(t) = V_{p} \nabla \cdot \boldsymbol{\tau} + m_p f_d\frac{\boldsymbol{u}_{f}(\boldsymbol{x}^i_p,t)-\boldsymbol{u}^i_{p}}{\tau_{p}}  + \boldsymbol{f}_{p}^{\mathrm{am},i} + \boldsymbol{f}_{p}^{\mathrm{lift}, i}. \label{eq:lpt_model}
 \end{equation}
 The first term on the right-hand side represents the effect of the undisturbed flow field \citep{maxeyEquationMotionSmall1983a}. The next term represents the drag force exerted on the particle. Here,
 $\tau_p=\rho_p d_p^2/(18\mu)$ is the particle response time and $f_d$ is an inertial drag correction. We use the one proposed by \citet{tavanashadParticleresolvedSimulationFreely2021} and derived from particle-resolved direct numerical simulations for particles with density ratio $\rho_p/\rho_f\leq 10$. This correction also accounts for volume fraction effects \citep{tavanashadParticleresolvedSimulationFreely2021}. The third term on the right-hand side of  (\ref{eq:lpt_model}) represents the added mass force and expresses as \citep{maxeyEquationMotionSmall1983a}
\begin{equation}
	\boldsymbol{f}_{p}^{\mathrm{am},i}=\frac{1}{2} \alpha_f \rho_f V_p \left ( \frac{d \boldsymbol{u}_p}{dt}-\frac{D \boldsymbol{u}_f(x^i_p,t)}{dt} \right).
\end{equation}
Lastly, $\boldsymbol{f}_p^{\mathrm{lift},i}$ is the Saffman lift force \citet{saffmanLiftSmallSphere1965}, which reads
\begin{equation}
	\boldsymbol{f}_p^{\mathrm{lift}} = 1.615 J \mu_{f} d_p |\boldsymbol{u}_{s}|\sqrt{\frac{d_p^2 |\boldsymbol{\omega}| \rho_f \alpha_f}{\mu_f}} \frac{\boldsymbol{\omega} \times \boldsymbol{u}_{s}}{|\boldsymbol{\omega}| |\boldsymbol{u}_{s}|}
\end{equation}
where $\boldsymbol{\omega} = \boldsymbol{\omega}_f(\boldsymbol{x^i_p},t)$ is the fluid vorticity at the particle location, $\boldsymbol{u}_{s} = \boldsymbol{u}_f(\boldsymbol{x}^i_p,t) - \boldsymbol{u}^i_p$ is the slip velocity, $J$ is a lift correction, which is equal to one in the model from \citet{saffmanLiftSmallSphere1965}. We use this model because \citet{costaInterfaceresolvedSimulationsSmall2020} showed that it best models the behavior of small inertial particles in comparison with PR-DNS simulations. 

We model collisions using the soft-sphere model described in \citet{capecelatroEulerLagrangeStrategy2013}. For the sake of brevity, we only give the highlights here and refer the reader to \citep{capecelatroEulerLagrangeStrategy2013} for further details. In brief, the force $\boldsymbol{f}_{p}^\mathrm{c,b\rightarrow a}$ exerted on particle $a$ due to collision with particle $b$ decomposes into normal and tangential components. We model the normal component $\boldsymbol{f}_{p,n}^\mathrm{c,b\rightarrow a}$ using a linearized spring-dashpot system, i.e,
\begin{equation}
	 \boldsymbol{f}_{p,n}^\mathrm{c,b\rightarrow a} = \begin{cases}
		-k\delta_{ab} \boldsymbol{n}_{ab}-\eta \boldsymbol{u}_{ab,n} & \text{if } |\boldsymbol{x}_{p}^a-\boldsymbol{x}_{p}^b| < 0.5(d_p^a+d_p^b)+\lambda \\
		0 & \mathrm{else}
	\end{cases} 
\end{equation}
where $\delta_{ab}= 0.5(d_p^a+d_p^b)-|\boldsymbol{x}_{p}^a-\boldsymbol{x}_{p}^b|$ is the the overlap between the two particles, $\boldsymbol{n}_{ab}$ is the unit normal vector between the two particles, and $\boldsymbol{u}_{ab,n}$ is the normal relative velocity. The parameters $k$ and $\eta$ denote the spring stiffness and damping factor, respectively. They relate to the reduced mass $m_{ab} = (1/m_a + 1/m_b)^{-1}$, collision time $\tau_{\mathrm{col}}$, and restitution coefficient $e$ as follows,
\begin{eqnarray}
	k &=& \frac{m_{ab}}{\tau_{\mathrm{col}}^2}(\pi^2+\mathrm{ln}(e)^2), \\
	\eta &=& -2 \mathrm{ln}(e) \frac{\sqrt{m_{ab}k}}{\sqrt{\pi^2+\mathrm{ln}(e)^2}}.
\end{eqnarray}
The so-called radius of influence $\lambda$ allows us to handle high-speed collisions robustly, by initiating the collision between high-speed pairs slightly before contact. Following \citep{finnParticleBasedModelling2016}, we compute $\lambda$ using
\begin{equation}
	\lambda = \frac{\lambda_0}{2}(d_{p,a}+d_{p,b})\left ( \frac{\mathrm{CFL}^c_{ab}}{\mathrm{CFL}^c_{\mathrm{max}}} \right )
\end{equation} 
where the collisional CFL number is $\mathrm{CFL}^c_{ab} = (2 |\boldsymbol{u}_{ab,n}| \Delta t)/(d_{p}^a+d_{p}^b)$
and $\lambda_0$ is the maximum radius of influence permitted when the collision occurs at the maximum collision CFL number, $\mathrm{CFL}^c_\mathrm{max}$. We model the tangential component of the collision force, with a static friction model
\begin{equation}
	\boldsymbol{f}_{p,t}^\mathrm{c,b\rightarrow a} = -\mu_s |\boldsymbol{f}_{p,t}^\mathrm{c,b\rightarrow a} | \boldsymbol{t}_{ab}
\end{equation}
where $\boldsymbol{t}_{ab}$ is the tangential direction and $\mu_s$ is the friction coefficient. 

We treat collisions with walls in the same manner as above, but considering that the wall has infinite mass.
Owing to the periodicity in the streamwise and spanwise directions, particles that cross the domain boundaries in these two directions wrap around to the other side of the domain.

In all simulations presented in this manuscript, the restitution coefficient is fixed at $e=0.65$, which is a first approximation of a restitution coefficient in a liquid medium. Nevertheless, we tested cases with restitution coefficient $e=0.9$ and found that the results are not sensitive to this parameter as shown in appendix \ref{sec:appendix_c}. Further, to ensure proper resolution of the contact, we use a stretched collision time $t_\mathrm{col}/\Delta t=15$ as described in \citep{capecelatroEulerLagrangeStrategy2013}. The friction coefficient is also fixed at $\mu_s=0.1$ in all runs.

The dynamics of the solid phase couple with those of the carrier phase via the momentum exchange field $\boldsymbol{F}_p$ and the volume fraction fields, $\alpha_p$ and $\alpha_f$. In the present volume-filtering framework, we compute these fields as follows
\begin{eqnarray}
	\boldsymbol{F}_{p}(\boldsymbol{x},t) &=& \sum_{i=1}^N \boldsymbol{f}^{h,i}_{p}(t) g(||\boldsymbol{x}-\boldsymbol{x}^i_{p}||)\label{eq:F_p}\\
	\alpha_{p}(\boldsymbol{x},t) &=& \sum_{i=1}^{N} V_{p} g(||\boldsymbol{x}-\boldsymbol{x}^i_{p}||), \\
	\alpha_{f}(\boldsymbol{x},t) &=& 1 - \alpha_{p}(\boldsymbol{x},t) 
\end{eqnarray}
where $g$ represents a Gaussian filter with width $\delta_{f} = 7 d_{p}$. 

Note that the governing equations (\ref{eq:continuity}) and (\ref{eq:momentum}) for the fluid phase are solved in both simulations with particles and without. In the latter case, $\alpha_f=1$ throughout the domain, which recovers the standard incompressible Navier-Stokes equations. Additional details on the solver, can be found in \citep{capecelatroEulerLagrangeStrategy2013}.
}

\subsection{Numerical experiments}
\label{sec:Num_experiments}

To understand the modulating effect of particles on the carrier turbulent flow, we perform Eulerian-Lagrangian simulations of a particle-laden channel flow at friction Reynolds $\Rey_\tau=u_{\tau}h/\nu=180$, where  $u_{\tau} = \sqrt{\tau_{w}/\rho}$ is the friction velocity and $\nu=\mu_f/\rho_f$ is the kinematic viscosity,  at increasing particle volume fraction. The presence of particles is expected to have a modulating effect on the turbulence which increases with channel-averaged particle volume fraction $\overline{\alpha}_p$. This is because the particles interact with the flow through at least three effects: (i) momentum exchange, via the term $\boldsymbol{F}_{p}$, (ii) volumetric displacement due to the volume occupied by the particles, and (iii) enhanced apparent viscosity. All these effects accentuate with increasing particle volume fraction, and may lead to flow statistics that differ considerably from those of a particle-free channel at the same Reynolds number. We compare the particle laden channels to a particle-free channel at $\Rey_{\tau} = 180$ to establish a baseline flow. Additionally, we compare to auxiliary particle-free simulations with viscosity matching the bulk apparent viscosity in the particle-laden cases. This is done to elucidate the effect of increasing apparent viscosity.

In addition to the friction Reynolds number, four other non-dimensional numbers control the dynamics of a particle-laden turbulent channel flow. These are the friction Stokes number $\Sto^{+}=\tau_{p} u_{\tau}^{2}/\nu$, the wall-scaled particle diameter $d_{p}^{+}=d_{p} u_{\tau}/\nu$, the particle to fluid density ratio $\rho_{p}/\rho_{f}$, and the channel mean particle volume fraction $\overline{\alpha}_p$.  Note that the superscript ``$^+$'' denotes normalization with inner wall units, i.e., $u_\tau$ for velocities, $\nu/u_\tau$ for lengths, and $\nu/u_{\tau}^2$ for time. In the following, we study the modulation of turbulence caused by increasing particle volume fraction solely. To isolate this effect, $\Sto^{+}$, $d_p^+$, and $\rho_p/\rho_f$ are fixed at about 7.9, 4.0, and 8.9 respectively. The volume fraction is varied by increasing the number of particles to yield values from 0.01 to 0.12. Varying the volume fraction also changes the mass loading $M= \rho_p \overline{\alpha}_p/(\rho_f \overline{\alpha}_f)$. This information is summarized in table \ref{tab:Phy_Param}.

\begin{table}
  \caption{Non-dimensional numbers in the present runs. Case 1 corresponds to a particle-free turbulent channel flow. Cases 2-5 are particle-laden flows where the particle volume-fraction is systematically increased from 1\% to 12\% by increasing the number of particles. \label{tab:Phy_Param}}
  \begin{ruledtabular}
    \begin{tabular}{l l l l l l l l l l}
      Parameter & Case 1 & Case 2 & Case 3 & Case 4 & Case 5 & Case 6 & Case 7 & Case 8 & Case 9 \\[1ex]
      $\Rey_\tau$ & 180 & 180 & 180 & 180 & 180 & 173.3 & 160.3 & 142.3 & 110.7\\
      $\overline{\alpha}_{p}$ & 0 & 0.01 & 0.03 & 0.06 & 0.12 & 0 & 0 & 0 & 0\\
      $M$         & 0 & 0.09 & 0.27 & 0.53 & 1.07 & 0 & 0 & 0 & 0\\
      $\rho_p/\rho_f$ & -- & 8.9 & 8.9 & 8.9 & 8.9 & -- & -- & -- & --\\
      $d_{p}^{+}$ & -- & 4 & 4 & 4 & 4 & -- & -- & -- & --\\
      $\Sto^{+}$ & -- & 7.9 & 7.9 & 7.9 & 7.9 & -- & -- & -- & --
    \end{tabular}
  \end{ruledtabular}
\end{table}

All five cases are simulated using the same domain size $L_{x} = 4 \pi h$, $L_{y} = 2h$, and $L_{z} = 4/3 \pi h$, where $h$ is the channel half height. The discretization is also the same in all cases and is 256 points in the stream-wise direction, 128 points in the wall-normal direction, and 168 points in the spanwise direction. The grid is spaced uniformly in the streamwise and spanwise directions. In the wall-normal direction, the grid is stretched using a hyperbolic tangent function such that the minimum spacing at the walls is $\Delta y^{+}_\mathrm{min} = 0.5$. The total number of Lagrangian particles varies from $1.17\times 10^5$, in case 2, up to $2.12\times 10^6$, in case 5.

\section{Results}

\subsection{Particle-free turbulent channel}
\label{sec:Stat_State}
We start by analyzing the results from the particle-free channel in case 1. Although this flow has been studied at length (see \citep{kimTurbulenceStatisticsFully1987}),  we present the main characteristics of this flow here in order to establish a baseline for comparison when particles are injected in the channel. To this end, we integrate equations (\ref{eq:continuity}) and (\ref{eq:momentum}) for a total of 60 eddy-turnover times, where we define the latter as $h/u_{\tau}$. The flow reaches a stationary state after about 40 eddy-turnover times. We use data from the last 20 eddy-turnover times, i.e., when turbulence is stationary, to compute the following statistics.

\begin{figure}
    \centering
	\begin{subfigure}{0.49\linewidth}
	\includegraphics[width = \linewidth]{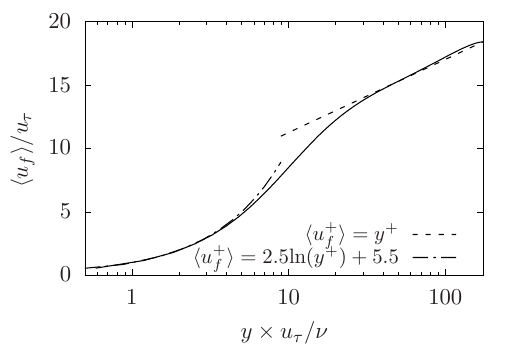}
	\caption{\label{fig:SP_stat_state_a}}
	\end{subfigure}
	\begin{subfigure}{0.49\linewidth}
	\includegraphics[width = \linewidth]{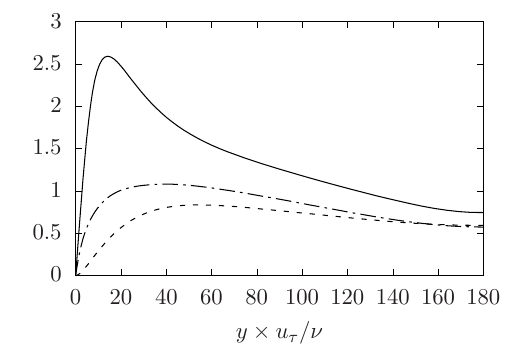}
	\caption{\label{fig:SP_stat_state_b}}
	\end{subfigure}
	\caption{
	Velocity statistics in a particle-free turbulent channel flow at $\Rey_\tau=180$. (a) Normalized mean streamwise velocity (\LineStyleSolid). (b) Normalized rms velocity fluctuations: $u'_\mathrm{rms}/u_\tau$ (\LineStyleSolid), $v'_\mathrm{rms}/u_\tau$ (\LineStyleDashed), and $w'_\mathrm{rms}/u_\tau$ (\LineStyleDashDot).
	\label{fig:SP_stat_state}}
\end{figure}

Figure \ref{fig:SP_stat_state} shows the mean streamwise velocity and the root-mean-square (rms) velocity fluctuations scaled by the friction velocity in the particle-free channel. The operation $\langle\cdot\rangle$ denotes ensemble and spatial averaging with respect to the streamwise and spanwise directions. The rms fluctuations  in the streamwise direction are defined as $u_{f,\mathrm{rms}}'=\langle u_f'^2\rangle^{1/2}$, where $u_f'=u_f-\left< u_f\right>$ is the streamwise velocity fluctuation. Fluctuations in the wall-normal and spanwise directions and their rms values are defined similarly. The mean streamwise velocity profile in figure (\ref{fig:SP_stat_state_a}) exhibits three layers commonly seen at this Reynolds number: a viscous layer for $0\lesssim y^+\lesssim 5$ where the profile follows the linear scaling $\langle u_f\rangle/u_\tau=y^+$, a buffer layer for $5\lesssim y^+\lesssim 30$, and a logarithmic layer for $y^+\gtrsim 30$ where the mean velocity profile follows $\langle u_f\rangle/u_\tau=2.5\mathrm{ln}(y^+)+5.5$. Figure \ref{fig:SP_stat_state_b} shows that the turbulent fluctuations are largest in the buffer layer and attenuate closer to the centerline. Turbulent fluctuations in the streamwise direction exceed those in the spanwise and wall-normal directions throughout the channel. The largest fluctuations are located at a wall-normal distance $y^+\sim 15$.

\begin{figure}
	\centering
    \includegraphics[width=5in,clip]{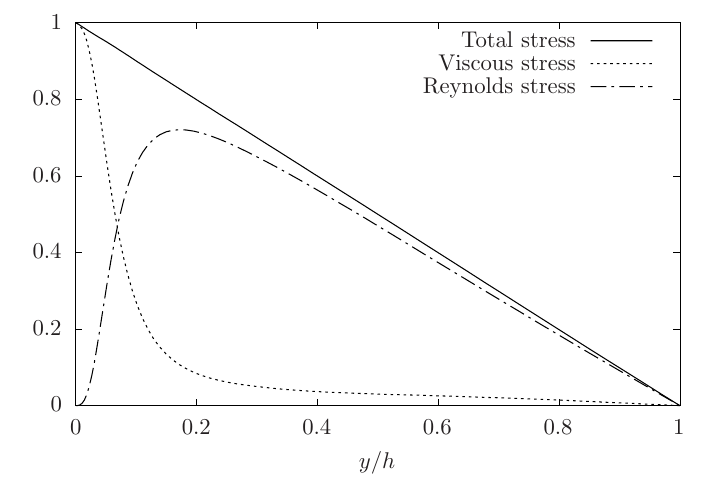}
    \caption{Stress balance in the particle-free channel. The total stress, sum of the viscous stress and Reynolds stresses, satisfies the linear relation (\ref{eq:stress_balance_SP}).}
    \label{fig:SP_Stress_Balance}
\end{figure}

For a particle-free channel, the balance between the viscous stress, pressure gradient, and Reynolds shear stress determines the structure of the flow. This balance can be obtained by averaging the momentum equation (\ref{eq:momentum}) and projecting in the streamwise direction. This procedure leads to
\begin{equation}
	\frac{d}{dy} \left( \mu_{f} \frac{d}{dy}\left< u_{f} \right> - \rho_{f} \left< u_{f}' v_{f}' \right> \right)= \left<\frac{\partial p}{\partial x}\right>   \label{eq:averaged_momentum_SP}
\end{equation}
where $\mu_{f} d \langle u_{f} \rangle /dy$ is the viscous stress, $-\rho_{f} \langle u_{f}' v_{f}' \rangle$ is the Reynolds shear stress, and $\langle \partial p/ \partial x \rangle= -A$ is the imposed pressure gradient, which is constant in our simulations. Integrating equation (\ref{eq:averaged_momentum_SP}) yields the well-known relationship
\begin{equation}
	\mu_{f} d \langle u_{f} \rangle /dy - \rho_{f} \langle u_{f}' v_{f}' \rangle = \tau_{w}(1-y/h) \label{eq:stress_balance_SP}
\end{equation}
where we have used the relationship $\tau_w/h= A$ that links the wall shear stress $\tau_{w} = \mu d \langle u_{f} \rangle / dy \rvert_{y=0}$ and pressure gradient. Equation (\ref{eq:stress_balance_SP}) shows that the total stress, due to viscous and Reynolds shear stresses, varies linearly across the channel. This is corroborated by figure \ref{fig:SP_Stress_Balance} which shows the total stress, Reynolds stress and viscous stress, normalized by the wall shear stress $\tau_{w}$ computed from the present simulations. As required by equation (\ref{eq:stress_balance_SP}), the total stress, sum of the viscous and Reynolds stresses, varies linearly across the channel. Close to the wall, the viscous stress dominates and the Reynolds stress is negligible. This follows from the no-slip and no-penetration conditions. However, the viscous stress drops quickly with wall-normal distance and the Reynolds stress in turn dominates.

\subsection{Particle-laden turbulent channel}
\label{sec:Volume_Frac}

\begin{figure}
    \centering
    \begin{subfigure}{0.75\linewidth}
      \begin{flushright}
        \includegraphics[width=2.0in]{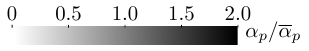}  
      \end{flushright}\vspace{-2ex}
      \includegraphics[width=\linewidth]{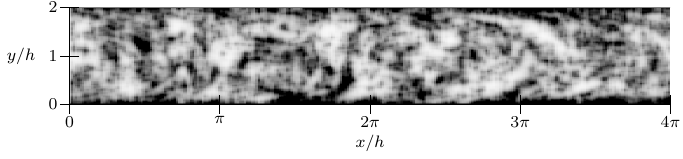}
      \caption{}
    \end{subfigure}
    \begin{subfigure}{0.75\linewidth}
      \includegraphics[width=\linewidth]{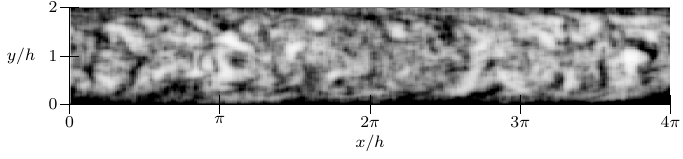}
      \caption{}
    \end{subfigure}
    \begin{subfigure}{0.75\linewidth}
      \includegraphics[width=\linewidth]{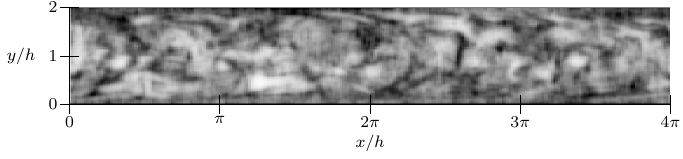}
      \caption{}
    \end{subfigure}
    \begin{subfigure}{0.75\linewidth}
      \includegraphics[width=\linewidth]{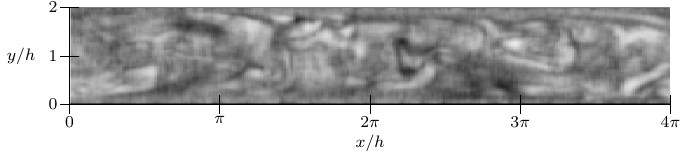}
      \caption{}
    \end{subfigure}
    \caption{Snapshot of the particle volume fraction field in a wall-normal plane at (a) $\overline{\alpha}_{p} = 1\%$, (b) $\overline{\alpha}_{p} = 3\%$, (c) $\overline{\alpha}_{p} = 6\%$, and (d) $\overline{\alpha}_{p} = 12\%$. Although the particle phase exhibits significant clustering at the lowest volume fraction, clustering reduces with increasing particle concentration.}
    \label{fig:Volume_Fraction_Visualizations}
\end{figure}

We now analyze cases 2 to 5, in-which solid particles are injected in the flow at increasing volume fraction from $1\%$ to $12\%$. To initialize these simulations, we first conduct particle-free simulations until the flow reaches a stationary state. Next, we inject particles into the domain with uniform distribution at the corresponding volume fraction in table \ref{tab:Phy_Param} for each case and with initial velocities equal to the fluid velocity at the particle locations. From there, we integrate the governing equations (\ref{eq:continuity}), (\ref{eq:momentum}), (\ref{eq:lpt_1}), and (\ref{eq:lpt_2}) up to a total of 140 eddy turnover time. Depending on the case, the flow reaches a new stationary state after about 20 eddy turnover time, for the channel at $\overline{\alpha}_p=1\%$, and 60 eddy turnover time, for the channel at $\overline{\alpha}_p=12\%$. We ignore data from the transient part and use only data from the stationary regime to compute flow and particle statistics.

\subsubsection{Qualitative analysis and leading order effects}

Although the particles are initially uniformly distributed, significant inhomogeneity develops by the time the particle-laden flow reaches a stationary state. Figure \ref{fig:Volume_Fraction_Visualizations} shows instantaneous snapshots of the normalized particle volume fraction field $\overline{\alpha}_p$ for the different suspensions considered in this study. We observe the formation of concentrated particle clusters and pockets of low volume fraction. Clustering is strongest at the low particle volume fraction $\overline{\alpha}_p=1\%$. In this case, the volume fraction within the clusters may exceed twice the channel average $\overline{\alpha}_p$. The low volume fraction pockets are also well pronounced, with certain regions dropping locally to $\overline{\alpha}_p=0$, i.e., pockets that are completely depleted of particles. However, the degree of clustering reduces considerably with increasing $\overline{\alpha}_p$, as the particles becomes progressively more evenly distributed. The case at $\overline{\alpha}_p=12\%$ shows the lowest level of clustering with local volume fraction not exceeding $\sim 1.2\times \overline{\alpha}_p$, although some inhomogeneity persists. 

\begin{figure}
  \begin{subfigure}{0.49\linewidth}
    \includegraphics[width=\linewidth]{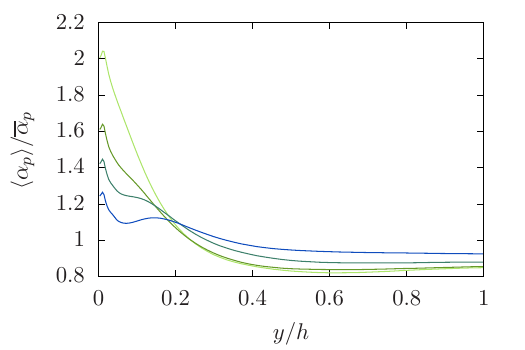}
    \caption{\label{fig:vfp_profile_a}}
  \end{subfigure}
  \begin{subfigure}{0.49\linewidth}
    \includegraphics[width=\linewidth]{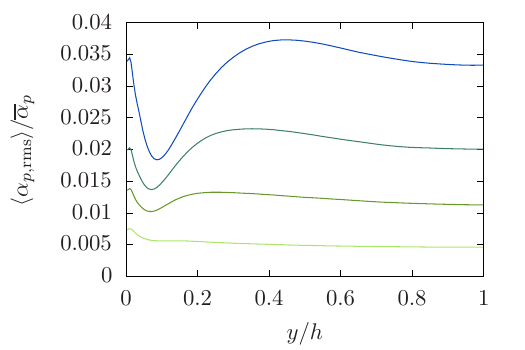}
    \caption{\label{fig:vfp_profile_b}}
  \end{subfigure}
  \caption{Variation of the particle volume fraction (a) mean and (b) rms fluctuations across the channel. \label{fig:vfp_profile}}
\end{figure}

For a more quantitative assessment of the particle distribution, we report in figure \ref{fig:vfp_profile} the profiles of particle volume fraction mean and rms fluctuations across the channel. In all cases the volume fraction profiles show some clustering at the wall with the greatest clustering occurring for $\overline{\alpha}_p = 1\%$. The near wall clustering decreases with increasing volume fraction from $2.2 \times \overline{\alpha}_p$ for $\overline{\alpha}_p = 1\%$ to $1.2 \times \overline{\alpha}_p$ for $\overline{\alpha}_p = 12\%$. Towards the channel center, the particle volume fraction plateaus at a value below the channel-averaged volume fraction due to the wall accumulation. At low volume fractions the high clustering in the near wall region may indicate a turbophoretic effect.  Figure \ref{fig:vfp_profile_b} shows the fluctuations in the fluid volume fraction. The fluctuations increase with increasing volume fraction, and the variation with the fluctuation profile increases as well.

\begin{figure}[h]
    \centering
    \begin{subfigure}{0.75\linewidth}
      \begin{flushright}
        \includegraphics[width=1.5in]{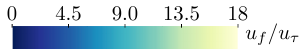}  
      \end{flushright}\vspace{-2ex}
      \includegraphics[width=\linewidth]{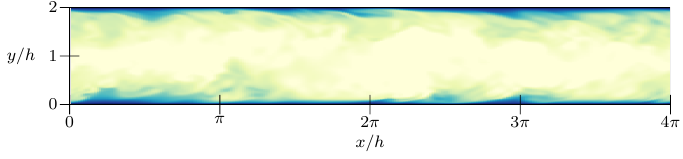}
      \caption{\label{fig:Fluid_Velocity_Visualizations_a}}
    \end{subfigure}    
    \begin{subfigure}{0.75\linewidth}
      \includegraphics[width=\linewidth]{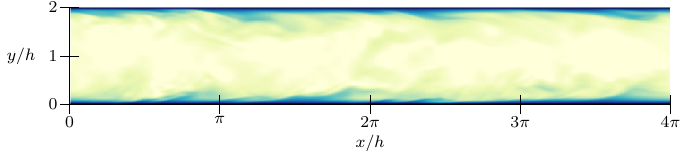}
      \caption{}
    \end{subfigure}
    \begin{subfigure}{0.75\linewidth}
      \includegraphics[width=\linewidth]{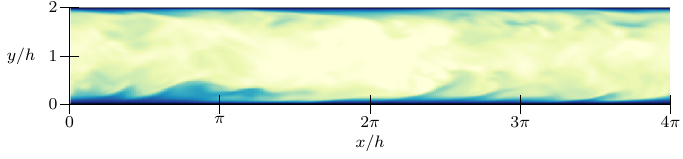}
      \caption{}
    \end{subfigure}
    \begin{subfigure}{0.75\linewidth}
      \includegraphics[width=\linewidth]{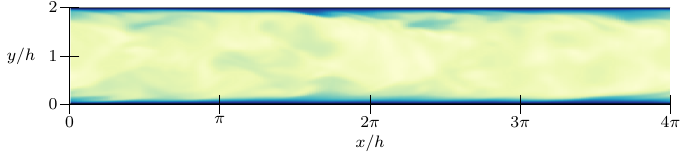}
      \caption{}
    \end{subfigure}
    \begin{subfigure}{0.75\linewidth}
      \includegraphics[width=\linewidth]{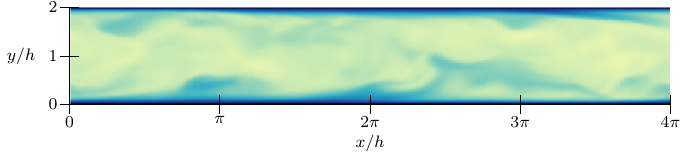}
      \caption{}
    \end{subfigure}
    \caption{Snapshot of the normalized streamwise fluid velocity in a wall-normal plane from (a) the particle-free channel ($\overline{\alpha}_p = 0\%$) and the particle-laden channels at (b) $\overline{\alpha}_p = 1\%$, (c) $\overline{\alpha}_p = 3\%$, (d) $\overline{\alpha}_p = 6\%$, and (e) $\overline{\alpha}_p = 12\%$. Turbulent fluctuations are suppressed with increasing particle volume fraction.\label{fig:Fluid_Velocity_Visualizations}}
\end{figure}

The snapshots of streamwise fluid velocity isocontours in figure \ref{fig:Fluid_Velocity_Visualizations} show that the suspended particles have a strong impact on the suspending turbulence. Compared to the particle-free case in figure \ref{fig:Fluid_Velocity_Visualizations_a}, the turbulent fluctuations appear to drop significantly with increasing $\overline{\alpha}_p$. This points to a laminarization of the flow field by the particles. The carrier flow velocity also appears to slow down in comparison with the particle-free channel. This suggests that the particles reduce the fluid mass flow rate through the channel considerably by reducing the bulk fluid velocity and through volume displacement.

\begin{figure}
  \centering
  \begin{subfigure}{0.49\linewidth}
    \includegraphics[width=\linewidth]{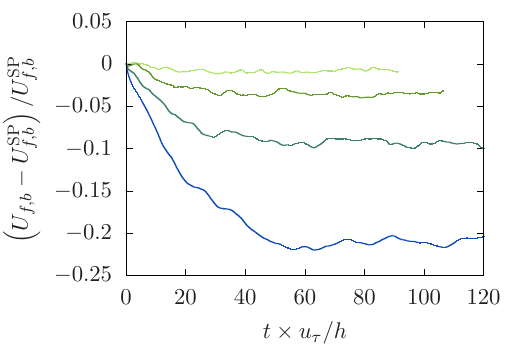}
    \caption{\label{fig:fluid_mass_flow_rate}}  
  \end{subfigure}
  \begin{subfigure}{0.49\linewidth}
    \includegraphics[width=\linewidth]{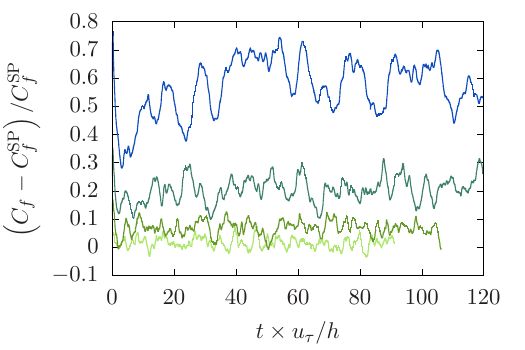}
    \caption{\label{fig:friction_coef}}  
  \end{subfigure}
  \caption{Relative change in fluid bulk velocity and friction coefficient at $\overline{\alpha}_p=1\%$ (\textcolor{gpltAlpha01}{\LineStyleSolid}), $\overline{\alpha}_p=3\%$ (\textcolor{gpltAlpha03}{\LineStyleSolid}), $\overline{\alpha}_p=6\%$ (\textcolor{gpltAlpha06}{\LineStyleSolid}), and $\overline{\alpha}_p=12\%$ (\textcolor{gpltAlpha12}{\LineStyleSolid}).
  \label{fig:mf_cf}}
\end{figure}

\begin{table}
  \caption{Summary of the effect of increasing the solid volume fraction on the fluid bulk velocity and friction coefficient.\label{tab:Results_table}}
  \begin{ruledtabular}
    \begin{tabular}{l l l l }
 
      Case & $\overline{\alpha}_p$ & $(U_{f,b}-U_{f,b}^\mathrm{SP})/U^\mathrm{SP}_{f,b}$ (\%)  & $(C_f-C_f^\mathrm{SP})/C_f^\mathrm{SP}$ (\%) \\[1ex]
         Case 2 & 0.01 & \textcolor{revision}{-0.77} & \textcolor{revision}{1.3}\\
         Case 3 & 0.03 & \textcolor{revision}{-3.6} & \textcolor{revision}{7.2}\\
         Case 4 & 0.06 & \textcolor{revision}{-9.3} & \textcolor{revision}{20.9}\\
         Case 5 & 0.12 & \textcolor{revision}{-21.0} & \textcolor{revision}{57.5}
    \end{tabular}
  \end{ruledtabular}
\end{table}

Figure \ref{fig:mf_cf} shows the variation of the relative fluid bulk velocity $U_{f,b}$, defined as 
\begin{equation}
  \label{eq:Bulk_Velocity_def}
  U_{f,b}=\frac{\dot{m}_f}{2hL_z\rho_f\overline{\alpha}_f}= \frac{\iint \rho_f\alpha_f u_fdydz}{2hL_z\rho_f\overline{\alpha}_f}=\frac{1}{2hL_z}\iint \frac{\alpha_f}{\overline{\alpha}_f} u_fdydz
\end{equation}
and friction coefficient over time. We use the friction velocity $u_{\tau}$, and friction coefficient $C_{f}^{\mathrm{SP}}$ from the unladen channel (case 1) for normalization. Note that, for the particle-laden channels, the friction coefficient is defined as
\begin{equation}
  C_f = \frac{\tau_w}{(1/2)\overline{\alpha}_f\rho_f U^2_{f,b}} \label{eq:def_cf}
\end{equation}
where $\overline{\alpha}_f=1-\overline{\alpha}_p$ is the channel-averaged fluid volume fraction.

The definition (\ref{eq:def_cf}) of $C_f$ for particle-laden cases is consistent with the standard definition for single-phase flows since the latter is recovered in the limit $\overline{\alpha}_p\rightarrow 0$. Further, definition (\ref{eq:def_cf}) can be interpreted as the friction coefficient of an effective single-phase fluid with apparent density $\rho^\star=\overline{\alpha}_f \rho_f$. It is also important to note that the wall shear stress $\tau_w$ varies, despite using the same forcing $A$ in all cases. This is because the two are related by the condition $\tau_w=\overline{\alpha}_{f} Ah$ (see appendix \ref{sec:appendix_a}). Consequently, $C_f$ can be written as
\begin{equation}
  C_f = \frac{A h}{(1/2) \rho_f U^2_{f,b}},
\end{equation}
which shows that variations in $C_f$ are inversely proportional to changes in bulk velocity.

Figure \ref{fig:fluid_mass_flow_rate} shows that the bulk velocity drops following the injection of particles in the domain. By the time the flow reaches a new stationary state, the bulk velocity $U_{f,b}$ levels off. Compared with the particle-free channel, the  bulk velocity drops by 2.2\%, 6.4\%, 14.4\%, and 30.2\%, in the channels at $\overline{\alpha}_p=1\%$, 3\%, 6\%, and 12\%, respectively. This is corroborated by the increase in relative friction coefficient shown in figure \ref{fig:friction_coef}. At $\overline{\alpha}_p=1\%$, 3\%, 6\%, and 12\%, the relative increase in skin-friction coefficient is $\Delta C_f/C_{f}^{\mathrm{SP}}= 0.05\%$, 3\%, 11\%, and 37\%, respectively. Table \ref{tab:Results_table} provides a summary of these values.

\subsubsection{Carrier flow velocity statistics}

\begin{figure}
    \begin{subfigure}{0.49\linewidth}
    \includegraphics[width=\linewidth]{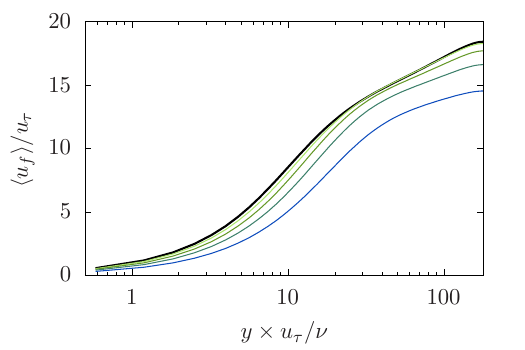}
    \caption{\label{fig:u_mean_profile}}
    \end{subfigure}
    \begin{subfigure}{0.49\linewidth}
    \includegraphics[width=\linewidth]{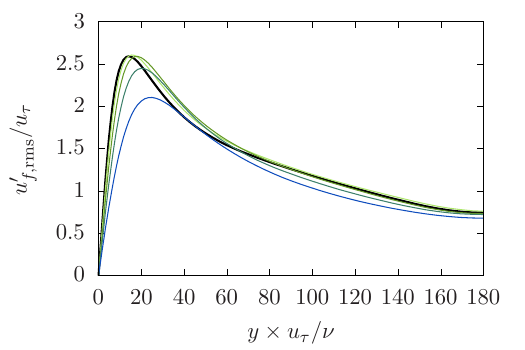}
    \caption{\label{fig:u_rms_profile}}
    \end{subfigure}
    \begin{subfigure}{0.49\linewidth}
    \includegraphics[width=\linewidth]{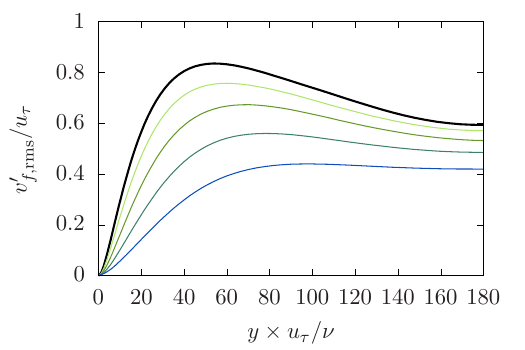}
    \caption{\label{fig:v_rms_profile}}
    \end{subfigure}
    \begin{subfigure}{0.49\linewidth}
    \includegraphics[width=\linewidth]{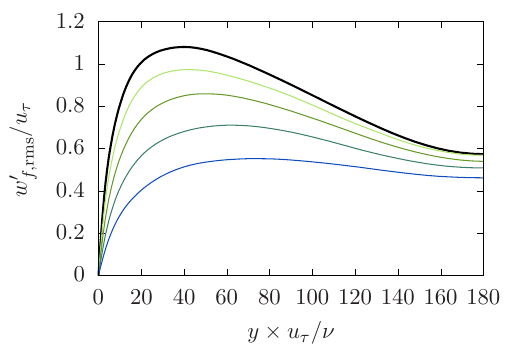}
    \caption{\label{fig:w_rms_profile}}
    \end{subfigure}
     \caption{Profiles of mean fluid velocity and rms fluctuations in the streamwise direction. The wall normal distance is scaled by the fixed friction Reynolds number $\Rey_{\tau} = 180$ The black lines (\textcolor{black}{\LineStyleSolidThick}) correspond to the particle-free ($\overline{\alpha}_p=0\%$) channel. Colored lines correspond to the particle-laden channels at $\overline{\alpha}_p=1\%$ (\textcolor{gpltAlpha01}{\LineStyleSolid}), $\overline{\alpha}_p=3\%$ (\textcolor{gpltAlpha03}{\LineStyleSolid}), $\overline{\alpha}_p=6\%$ (\textcolor{gpltAlpha06}{\LineStyleSolid}), and $\overline{\alpha}_p=12\%$ (\textcolor{gpltAlpha12}{\LineStyleSolid}). Increasing the particle concentration tends to slow down the carrier flow and suppress turbulent fluctuations.}
\end{figure}

To understand how turbulence in the carrier flow is modified, we start by analyzing changes to the fluid velocity statistics.

Figure \ref{fig:u_mean_profile} shows the effect of increasing mean particle volume fraction $\overline{\alpha}_p$ on the mean streamwise velocity profile. In all four cases, the particles reduce the streamwise velocity throughout the channel. As the volume fraction increases, the velocity is further reduced. The logarithmic layer, indicating turbulence, is present in cases 2-4, but is shortened with increasing volume fraction. This indicates progressive relaminarization of the flow structures with increasing $\overline{\alpha}_p$. This further supports the qualitative observations from figure \ref{fig:Fluid_Velocity_Visualizations} which showed a decrease in fluid velocity coupled with suppression of fluid structures. Additionally, the reduction of fluid velocity throughout the channel directly reduces the fluid bulk velocity and increases coefficient of friction, as observed previously.

Figure \ref{fig:u_rms_profile} shows profiles of the streamwise rms velocity fluctuations. For $\overline{\alpha}_{p} = 1\%$ and 3\%, the peak streamwise fluctuation does not change significantly, while in cases $\overline{\alpha}_{p} = 6\%$, and 12\% it is decreased by 5.30\% and 19.5\%, respectively. This peak is shifted towards the channel center in each case, with the shift increasing with increasing volume fraction. While it occurs at $y^{+} = 12$ in the particle free-case 1, the peak is shifted towards $y^{+} \simeq 16$, 17, 20, and 24 in the particle-laden cases with $\overline{\alpha}_{p} = 1\%$, 3\%, 6\%, and 12\%, respectively.

Increasing volume fraction reduces the wall-normal and cross-stream velocity fluctuations as shown in figures \ref{fig:v_rms_profile} and \ref{fig:w_rms_profile}. The reduction is most notable at volume fraction $\overline{\alpha}_{p} = 12\%$ which shows a maximum decrease of $46.5\%$ for the wall-normal fluctuations and $47.5\%$ in the cross-stream fluctuations. As with the streamwise fluctuations, the peaks values are shifted towards the center of the channel, with the shift increasing with increasing volume fraction. The suppression of velocity fluctuations further indicates that turbulence has been reduced.

\subsubsection{Modified stress balance}

\begin{figure}
  \centering
    \includegraphics[width=5in,clip]{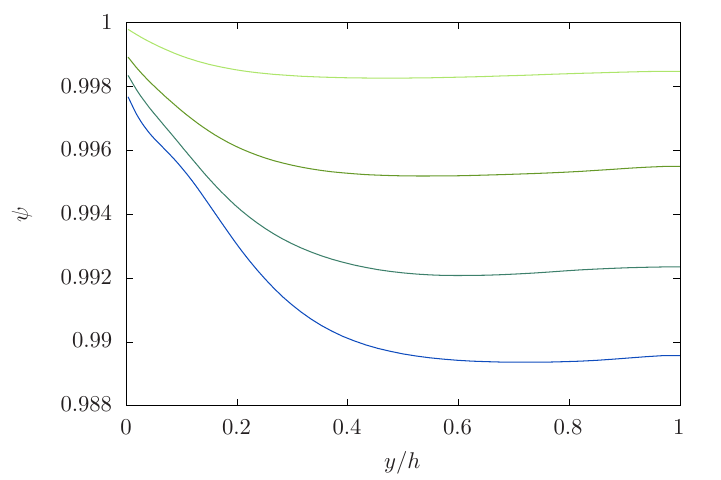}
  \caption{Stress distortion variation with increasing volume fraction $\overline{\alpha}_p=1\%$ (\textcolor{gpltAlpha01}{\LineStyleSolid}), $\overline{\alpha}_p=3\%$ (\textcolor{gpltAlpha03}{\LineStyleSolid}), $\overline{\alpha}_p=6\%$ (\textcolor{gpltAlpha06}{\LineStyleSolid}), and $\overline{\alpha}_p=12\%$ (\textcolor{gpltAlpha12}{\LineStyleSolid}).
  \label{fig:Psi_variation}}
\end{figure}

Introducing particles in the channel leads to an additional contribution to the stress budget. Similar to the analysis shown in \S\ref{sec:Stat_State},  we obtain the modified stress balance by Reynolds-averaging the fluid-phase momentum equations. Details of this procedure can be found in appendix A. The resulting stress balance is
\begin{equation}
	\langle \mu_{f}^\star  \frac{\partial}{\partial y} u_{f} \rangle-\rho_f \langle \alpha_{f} u_{f}''v_{f}'' \rangle + \langle \Tau_{p} \rangle = \tau_{w} \psi \left( 1-\frac{y}{h} \right) \label{eq:total_stress}
\end{equation}
where $\langle\Tau_{p}\rangle$ is the ensemble-averaged particle stress obtained from the particle feedback force (\ref{eq:F_p}) as follows
\begin{equation}
		\langle \Tau_{p} \rangle = \int_0^{y} \langle F_{p,x} \rangle dy' - \frac{1}{2h} \int_0^{2h} F_{p,x} dy
\end{equation}
\textcolor{revision}{The double primes indicate fluctuations with respect to Favre averaging, that is
\begin{equation}
	\widetilde{u}_f = \frac{\langle \alpha_p u_f \rangle}{\langle \alpha_p \rangle},
\end{equation}
which allows the decomposition of the instantaneous velocity field as
\begin{equation}
	u_f = \widetilde{u}_f + u_f''.
\end{equation}}
The quantity $\psi$ is a ``stress distortion term". It expresses as
\begin{equation}
	\psi = \left( 1-\frac{1}{h}\int_{0}^{y} \frac{\langle \alpha_{f} \rangle}{\overline{\alpha}_f} dy' \right)/\left( 1-\frac{y}{h} \right)
\end{equation}
If $\psi\ne 1$, the total stress (right hand side of equation (\ref{eq:total_stress})) could display an unusual non-linear variation with wall-normal distance. However, for this to be the case, very high particle volume fractions and very strong particle clustering are required simultaneously.  This is not the case in our present simulations, as the relative variations of fluid volume fraction $\langle \alpha_{f} \rangle/\overline{\alpha}_f$ remain small despite significant particle clustering. Figure \ref{fig:Psi_variation} shows the variation of $\psi$ across the channel for the particle-laden cases we have considered. The deviation from unity does not exceed 1 percent in any case. Importantly, this means that the total stress for the particle-laden cases herein must follow the same linear behavior as in a particle-free channel. 

\begin{figure}
  \begin{subfigure}{0.49\linewidth}
    \includegraphics[width=\linewidth]{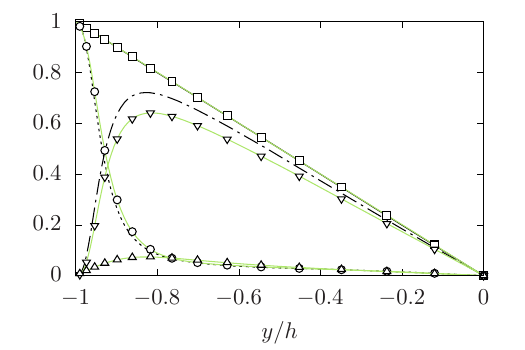}
    \caption{$\overline{\alpha}_p = 1\%$}
  \end{subfigure}
  \begin{subfigure}{0.49\linewidth}
    \includegraphics[width=\linewidth]{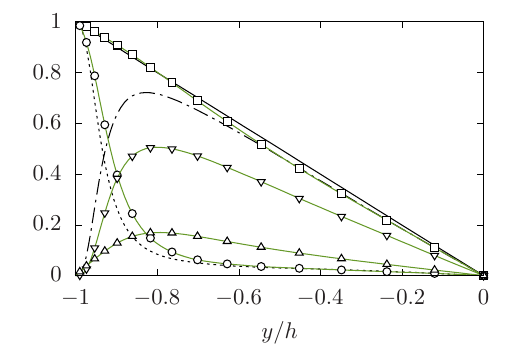}
    \caption{$\overline{\alpha}_p= 3\%$}
  \end{subfigure}
  \begin{subfigure}{0.49\linewidth}
    \includegraphics[width=\linewidth]{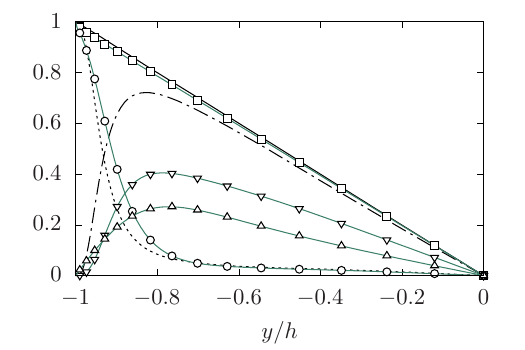}
    \caption{$\overline{\alpha}_p= 6\%$}
  \end{subfigure}
  \begin{subfigure}{0.49\linewidth}
    \includegraphics[width=\linewidth]{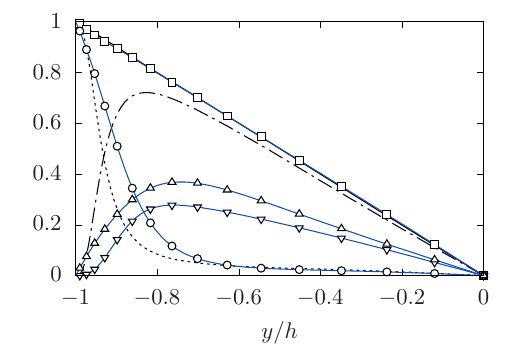}
    \caption{$\overline{\alpha}_p = 12\%$}
  \end{subfigure}
  \caption{Balance of stresses in particle-laden channels at solid volume fraction (a) $\overline{\alpha}_p = 1\%$, (b) $\overline{\alpha}_p = 3\%$, (c) $\overline{\alpha}_p = 6\%$, and (d) $\overline{\alpha}_p = 12\%$. Black lines correspond to data from the particle-free channel as in figure \ref{fig:SP_Stress_Balance}. Lines with symbols correspond to data from the particle-laden channels: total stress (\LineStyleSolidSquare), viscous stress (\LineStyleSolidCircle), Reynolds stress (\LineStyleSolidTriangleD), and particle stress (\LineStyleSolidTriangleU). Increasing volume fraction leads to increased particle stress and decreased Reynolds stress. The decrease in Reynolds stress causes an expansion of the viscous stresses towards the channel center. At $\overline{\alpha}_p = 12\%$, the particle stress is greater than the Reynolds stress.
    \label{fig:Stress_Balances}}
\end{figure}

The sum of these stresses, along with each stress component, is plotted in figure \ref{fig:Stress_Balances}. In all cases, the expected linear behavior found through the stress analysis is obeyed. The particle stress $\langle T_p\rangle$ has the smallest magnitude in case 2, where $\overline{\alpha}_p=1\%$ is the lowest. In this case, $\langle T_p\rangle$ contributes to the reduction of the Reynolds stress by a proportional  amount and the expansion of the viscous stress towards the center of the channel. The latter suggests a slight expansion of the viscous layer, as previously observed in figure \ref{fig:u_mean_profile_Mu_ef}. With increasing particle volume fraction, the particle stress increases while the Reynolds shear stress decrease, until the particle stress exceeds the Reynolds stress, as shown in case 5. At the same time, the viscous stress continues expanding towards the channel center with increasing volume fraction. This further indicates that the flow relaminarizes due to the presence of particles. \textcolor{revision}{\citet{picanoTurbulentChannelFlow2015a,lashgariTurbulentChannelFlow2017,costaNearwallTurbulenceModulation2021} performed particle-resolved direct numerical simulations of particle-laden turbulent channel flow. They similarly showed the presence of particle stresses which modify the overall stress balance by altering the viscous and reynolds stresses. However, differences in particle size, density, and volume fraction mean that these represent a different regime of modulation.}

\subsection{Mechanisms of particle-flow interaction}
To understand the reasons behind the relaminarization of the flow and increase in skin-friction drag with increasing particle volume fraction, it is necessary to detangle two effects caused by the suspended particle: (1) the increasing apparent viscosity of the suspension and (2) the turbulence modulation by the particle feedback force.

From expression (\ref{eq:def_cf}), we see that the increase in skin-friction drag coefficient with  $\overline{\alpha}_p$ results from a corresponding drop in the fluid bulk velocity $U_{b,f}$, as the product of the pressure gradient with the channel half height $Ah$ is the same in all  configurations presented here. In the case of a particle-free channel, the bulk velocity relates to the stresses in a particle-free channel following \citep{daveMechanismsDragReduction2023}
\begin{equation}
	\frac{U^{SP}_{f,b}}{u_{\tau}} = \frac{\Rey_{\tau}}{3}\left ( 1  + \frac{3}{(u_{\tau} h)^2}\int^h_0 \int_0^y \langle u'_f v'_f \rangle dy'dy \right)
	 \label{eq:bulk_velocity_expression_SP}
\end{equation}
which shows that the Reynolds shear stress causes a reduction of the bulk velocity, since $ \langle u'_f v'_f \rangle$ is negatively signed, compared to the laminar case ($\langle u'_f v'_f \rangle=0$). However, the presence of inertial particles modifies $U_{f,b}$. As we show in appendix \ref{sec:appendix_b}, the bulk velocity in a slurry relates to the stresses in the channel following
\begin{equation}
	\frac{U_{f,b}}{u_{\tau}} = \frac{\Rey^\star_{\tau}}{3}\left ( 1  + \frac{3}{(u_{\tau} h)^2}\int^h_0 \int_0^y \left( \langle u''_f v''_f \rangle - M \frac{\langle \Tau_p \rangle}{ \rho_p\overline{\alpha}_p } \right)dy'dy \right)
	 \label{eq:bulk_velocity_expression}
\end{equation}
where $\Rey^\star_{\tau}$ is the suspensions' effective friction Reynolds number
\begin{equation}
	\Rey^\star_{\tau} = u_{\tau} h /\nu^{\star} = \overline{\alpha}_f^{3.8} \Rey_{\tau},
\end{equation}
where $\nu^{\star} = \mu^{\star}/\rho^{\star}$ is the effective kinematic viscosity, $\mu^{\star} = \alpha_f^{-2.8} \mu_f$ is the effective dynamic viscosity, and $\rho^{\star} = \alpha_f \rho_f$ is the effective density. Note that, in our simulations with constant-pressure gradient forcing, the friction velocity is the same in all cases as
\begin{equation}
	u_{\tau} = \sqrt{\tau_w/ \overline{\alpha}_f \rho_{f}}=\sqrt{\overline{\alpha}_{f} A h/ \overline{\alpha}_f \rho_{f}}=\sqrt{A h/ \rho_{f}}.
\end{equation}
Expression (\ref{eq:bulk_velocity_expression}) shows that turbulence suppression by the particles is due to both the increased viscosity and possibly interaction of particles with turbulence scales. Indeed, the role of increased viscosity appears to be the dominant effect as seen from the decreased bulk velocity. 

To detangle these two effects, we present comparisons with additional particle-free simulations, cases 6-9, where we match the fluid kinematic viscosity to the apparent kinematic viscosity of the suspension for each case $\overline{\alpha}_{f} = 1\%$, 3\%, 6\%, and 12\%. This leads to flows with friction Reynolds numbers that matches the effective friction Reynolds number $\Rey_\tau^\star$ from the slurry cases. Comparing with these flows allows us to separate the modulation of turbulent fluctuations due to particles from their effect on the suspensions' effective viscosity. Table \ref{tab:aux_Phy_Param} shows the parameters for these additional particle-free channel flow simulations. 
\begin{table}
  \caption{Comparison between the particle-laden and particle-free channels with matching apparent kinematic viscosity. \label{tab:aux_Phy_Param}}
  \begin{ruledtabular}
    \begin{tabular}{l l l l}
       $\Rey^{\star}_\tau$ & $\overline{\alpha}_p$  & $(U_{f,b}-U_{f,b}^\star)/U^{\star}_{f,b}$ (\%)  & $(C_f-C_f^\star)/C_f^{\star}$ (\%) \\[1ex]
        173 & 0.01 & \textcolor{revision}{-0.4}  & \textcolor{revision}{0.6} \\
	    160 & 0.03 & \textcolor{revision}{-2.3}  & \textcolor{revision}{4.2} \\
        142 & 0.06 & \textcolor{revision}{-7.3}  & \textcolor{revision}{15.4} \\
        110 & 0.12 & \textcolor{revision}{-16.6} & \textcolor{revision}{40.9} 
    \end{tabular}
  \end{ruledtabular}
\end{table}

\begin{figure}
    \begin{subfigure}{0.49\linewidth}
    \includegraphics[width=\linewidth]{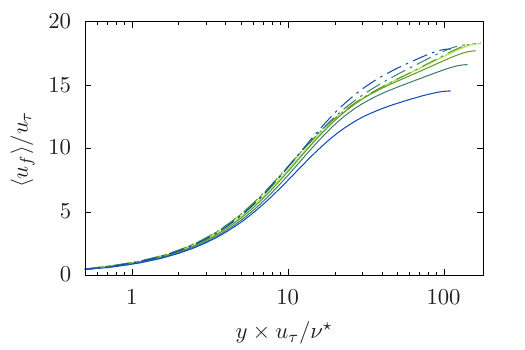}
    \caption{\label{fig:u_mean_profile_Mu_ef}}
    \end{subfigure}
    \begin{subfigure}{0.49\linewidth}
    \includegraphics[width=\linewidth]{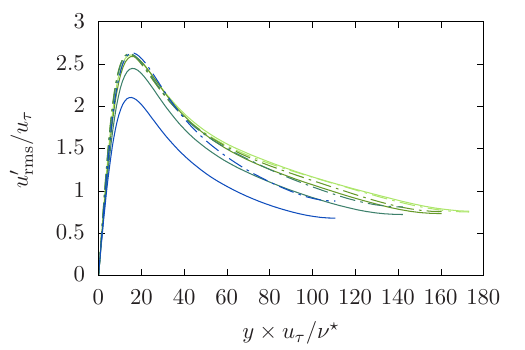}
    \caption{\label{fig:u_rms_profile_Mu_ef}}
    \end{subfigure}
    \begin{subfigure}{0.49\linewidth}
    \includegraphics[width=\linewidth]{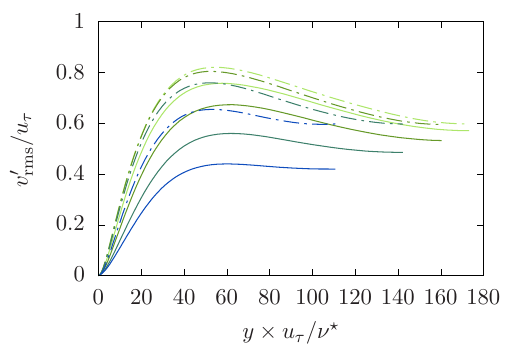}
    \caption{\label{fig:v_rms_profile_Mu_ef}}
    \end{subfigure}
    \begin{subfigure}{0.49\linewidth}
    \includegraphics[width=\linewidth]{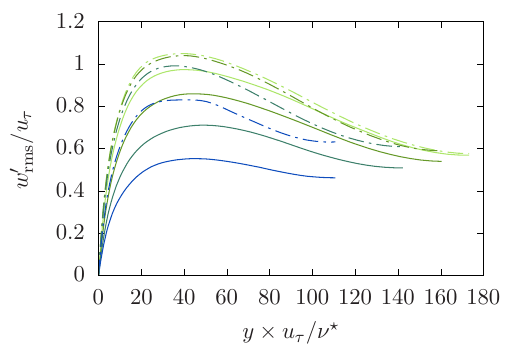}
    \caption{\label{fig:w_rms_profile_Mu_ef}}
    \end{subfigure}
     \caption{Profiles of mean fluid velocity and rms velocity fluctuations. Here, the wall-normal position is scaled using the friction Reynolds number and the apparent kinematic viscosity. Solid lines correspond to the particle-laden channels, while dash-dotted lines correspond to the auxiliary particle-free channels with matching apparent kinematic viscosity at $\overline{\alpha}_p=1\%$ (\textcolor{gpltAlpha01}{\LineStyleSolid}/\textcolor{gpltAlpha01}{\LineStyleDashDot}), $\overline{\alpha}_p=3\%$ (\textcolor{gpltAlpha03}{\LineStyleSolid}/\textcolor{gpltAlpha03}{\LineStyleDashDot}), $\overline{\alpha}_p=6\%$ (\textcolor{gpltAlpha06}{\LineStyleSolid}/\textcolor{gpltAlpha06}{\LineStyleDashDot}), and $\overline{\alpha}_p=12\%$ (\textcolor{gpltAlpha12}{\LineStyleSolid}/\textcolor{gpltAlpha12}{\LineStyleDashDot}).  Increasing the particle concentration tends to slow down the carrier flow and suppress turbulent fluctuations further than one would expect solely based on the enhanced apparent viscosity.\label{fig:profile_Mu_ef}}
\end{figure}

Comparison with the particle-free channels at matching $\Rey_\tau^\star$ shows that  solid particles slow down the carrier flow and suppress turbulent fluctuations further than one would expect solely based on the enhanced apparent viscosity. Figure \ref{fig:u_mean_profile_Mu_ef} shows the streamwise velocity profile for each particle-laden case alongside its  corresponding particle-free case at matching effective friction Reynolds number $\Rey_\tau^\star$. For the latter cases, the profiles show similar values in the viscous and buffer layers, while in the logarithmic region the streamwise velocity is slightly higher in the particle-free cases with increased viscosity. However, the logarithmic region is shortened with decreasing $\Rey_\tau^\star$. This, unsurprisingly, shows a decrease in turbulence with decreasing Reynolds number. The decrease in the streamwise velocity in the particle-free cases is much less than the decrease in the particle-laden cases. The presence of particles leads to a greater reduction in the streamwise velocity than the decrease from increased viscosity alone. 
Next, we consider the velocity fluctuations, which are shown in figures \ref{fig:u_rms_profile_Mu_ef}, \ref{fig:v_rms_profile_Mu_ef}, and \ref{fig:w_rms_profile_Mu_ef}. When the wall-normal position is normalized with the effective inner wall scale $\nu^\star/u_\tau$, the streamwise velocity fluctuation peaks at the same location $y u_\tau/\nu^\star\sim 15$ in both particle-laden and particle-free channels with matching $\Rey_\tau^\star$. At solid volume fractions $\overline{\alpha}_p = 1\%$ and 3\%, the magnitude of the streamwise velocity fluctuations match in the particle-laden and corresponding particle-free channels. However, the streamwise velocity fluctuations are significantly lower in the particle-laden channels at $\overline{\alpha}_p = 6\%$ and 12\% compared to the particle-free channels at the same $\Rey^\tau$.
Further, the wall-normal and spanwise fluctuations present significant discrepancy between  the particle-laden and particle-free channels at corresponding $\Rey_\tau^\star$. This discrepancy increases with increasing solid volume fraction.

\begin{figure}
  \begin{subfigure}{0.49\linewidth}
    \includegraphics[width=\linewidth]{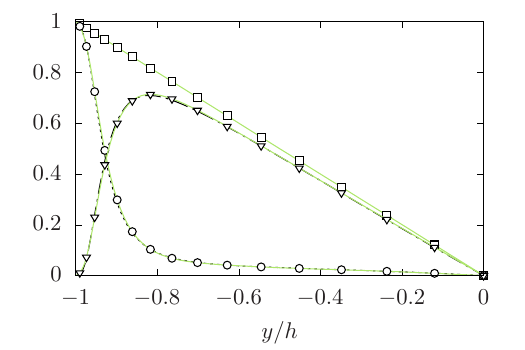}
    \caption{$\overline{\alpha}_p = 1\%$}
  \end{subfigure}
  \begin{subfigure}{0.49\linewidth}
    \includegraphics[width=\linewidth]{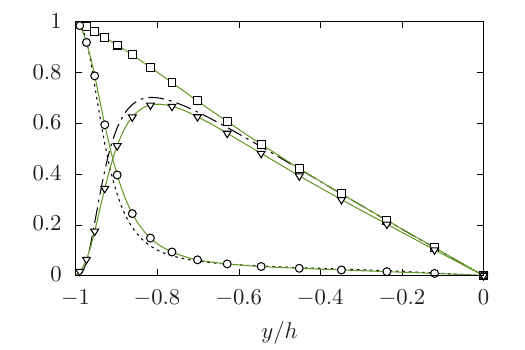}
    \caption{$\overline{\alpha}_p= 3\%$}
  \end{subfigure}
  \begin{subfigure}{0.49\linewidth}
    \includegraphics[width=\linewidth]{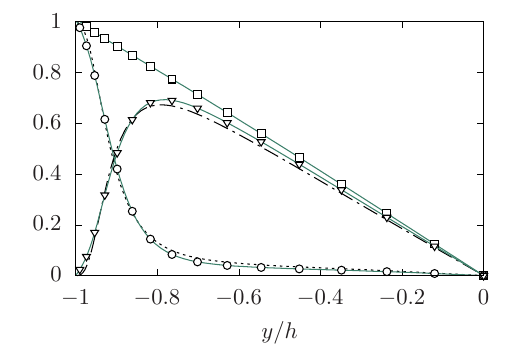}
    \caption{$\overline{\alpha}_p= 6\%$}
  \end{subfigure}
  \begin{subfigure}{0.49\linewidth}
    \includegraphics[width=\linewidth]{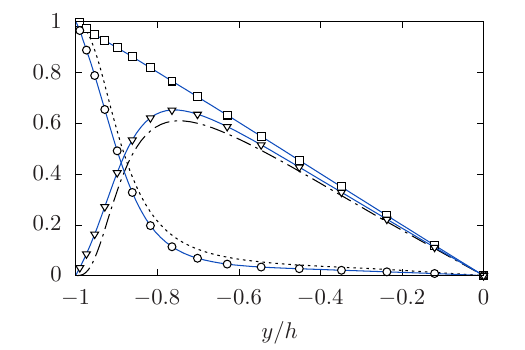}
    \caption{$\overline{\alpha}_p = 12\%$}
  \end{subfigure}
  \caption{
  Comparison of the stresses in particle-laden (lines with symbols) and particle-free channels (black lines) with matching friction Reynolds numbers $\Rey_\tau^\star$ for solid volume fractions (a) $\overline{\alpha}_p = 1\%$, (b) $\overline{\alpha}_p = 3\%$, (c) $\overline{\alpha}_p = 6\%$, and (d) $\overline{\alpha}_p = 12\%$.  Total stress (\LineStyleSolid/\LineStyleSolidSquare), viscous stress (\LineStyleDashed/\LineStyleSolidCircle), Reynolds stress in particle-free channels (\LineStyleDashDot), and sum of Reynolds and particle stresses in particle-laden channels (\LineStyleSolidTriangleD). The discrepancy between stresses in the particle-laden and particle-free channels with matching $\Rey_\tau^\star$ increases with solid volume fraction.    \label{fig:Stress_Balances_Mu_ef}
    }
\end{figure}

With increasing $\overline{\alpha}_p$, the increasing discrepancy between stresses in the particle-laden and particle-free channels with matching $\Rey_\tau^\star$ shows that the particle feedback force plays a growing role in turbulence modulation. At the solid volume fraction $\alpha_p=1\%$ and $\alpha_p=3\%$, figure \ref{fig:Stress_Balances_Mu_ef} shows that stresses in the particle-laden and particle-free channels at matching $\Rey_\tau^\star$ present good agreement. Here, it is imperative to compare the Reynolds stress in the particle-free channel to the sum of the Reynolds and particle stresses in the particle-laden channel. Starting with $\alpha_p=6\%$, the viscous stress in the particle-laden channels contracts as the sum of the Reynolds and particle stresses grows. These effects accentuate at $\alpha_p=12\%$.

\begin{figure}
  \centering
    \includegraphics[width=5in,clip]{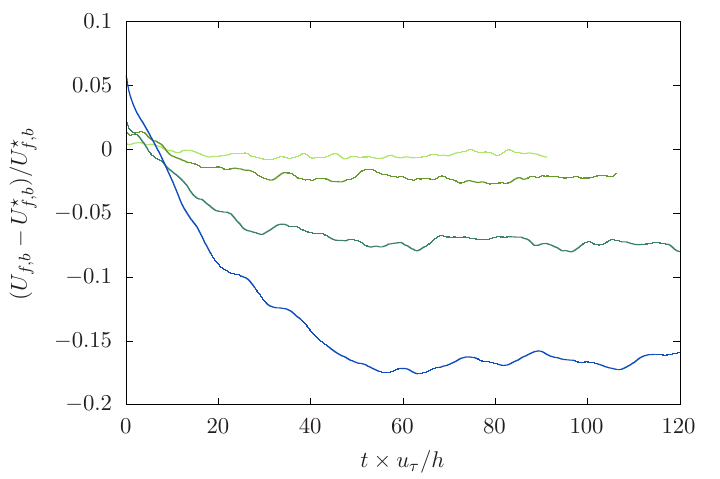}
  \caption{Relative change in bulk velocity and friction coefficient at $\overline{\alpha}_p=1\%$ (\textcolor{gpltAlpha01}{\LineStyleSolid}), $\overline{\alpha}_p=3\%$ (\textcolor{gpltAlpha03}{\LineStyleSolid}), $\overline{\alpha}_p=6\%$ (\textcolor{gpltAlpha06}{\LineStyleSolid}), and $\overline{\alpha}_p=12\%$ (\textcolor{gpltAlpha12}{\LineStyleSolid}).
  \label{fig:fluid_mass_flow_rate_Mu_ef}}
\end{figure}

In addition to modifying the stress balance, the particle feedback force has a large impact on the fluid mass transport in the denser channels.
Using equation (\ref{eq:bulk_velocity_expression}), we can express the normalized change in bulk velocity as
\begin{equation}
	\frac{1}{u_{\tau}}(U_{f,b}-U^{\star}_{f,b})= \frac{\Rey^{\star}_{\tau}}{(u_{\tau}h)^2} \int^h_0 \int^y_0
	\left(\langle  u''_f v''_f\rangle + M \frac{\langle \Tau_p \rangle}{\rho_p \overline{\alpha}_p}\right)
	-
	\langle  {u_f^{\star}}' {v_f^{\star}}' \rangle dy' dy 
\end{equation}
where the superscript $\star$ denotes quantities from the particle-free channels at the appropriate  $\Rey_\tau^\star$. From this expression, it is clear that if the sum of the Reynolds and particle stresses for the particle-laden channel is greater than the Reynolds stress for the particle-free channel, then the bulk fluid velocity will decrease. To verify this we plot the relative change in bulk fluid velocity in figure \ref{fig:fluid_mass_flow_rate_Mu_ef}. As previously noted with the stresses, there is little difference between the particle-free channels at $\Rey_\tau^\star=173$ and 160 and the particle-laden channels at $\alpha_p=1\%$ and 3\%, respectively. For these two cases, the relative change in fluid bulk is -0.8\% and -1.9\% (see table \ref{tab:aux_Phy_Param}). Moreover, the relative change in friction coefficient is 0.2\% and 0.5\%. The greater discrepancy between the stresses in the channels at $\alpha_p=6\%$ and $\alpha_p=12\%$ and their corresponding particle-free channels with matching apparent kinematic viscosity leads to greater differences in the bulk velocity and friction coefficient. At $\alpha_p=6\%$, $(U_{f,b}-U_{f,b}^\star)/U^{\star}_{f,b}= -6.5\%$  and  $(C_f-C_f^\star)/C_f^{\star}= 7.0\%$, while at $\alpha_p=12\%$, $(U_{f,b}-U_{f,b}^\star)/U^{\star}_{f,b}= -16.1\%$  and  $(C_f-C_f^\star)/C_f^{\star}= 25.2\%$.

\section{Conclusions}
\label{sec:conclus}
We investigate the mechanisms by which inertial particles modulate turbulence in wall-bounded turbulent slurries. Using Euler-Lagrange simulations of turbulent liquid-solid channel flow at friction Reynolds numbers $\Rey_{\tau} = 180$, friction Stokes number $\Sto^+=7.9$, density ratio $\rho_p/\rho_f=8.9$, and solid volume fraction ranging from $\overline{\alpha}_p=1\%$ up to $\overline{\alpha}_p=12\%$, we show that the particles have a large impact on mass transport and turbulence in the carrier fluid. We show that the mechanisms underpinning the flow modulation are two fold: (I) the increase of the suspension's apparent kinematic viscosity with increasing solid volume fraction and (II) turbulence modulation through the particle feedback force. The first mechanism accounts for most of the flow changes we observe in channels with $\overline{\alpha}_p \leq 3\%$, namely, the reduction of turbulent fluctuations and increase in friction coefficient. In denser channels, the particle feedback force plays a large role and contributes to further reduction of turbulence and increase in friction coefficient.

In comparison with single-phase turbulent channel flow at $\Rey_{\tau} = 180$, dispersing solid particles modifies the carrier fluid's mass and momentum transport considerably. The mean streamwise fluid velocity decreases with increasing solid volume fraction. Likewise,  velocity fluctuations in the streamwise, wall-normal, and spanwise directions decrease with increasing solid volume fraction. In the present simulations, where the pressure gradient is maintained constant, the bulk fluid velocity reduces by -1\% at $\overline{\alpha}_p =1\%$ and by -20.6\% at $\overline{\alpha}_p =12\%$. The friction coefficient increases by 2\% at $\overline{\alpha}_p =1\%$ and up to 59\% at $\overline{\alpha}_p = 12\%$. The structure of the flow also changes as the particles modify the stress balance in the channel. With increasing solid volume fraction, the contribution of the viscous stress to the stress balance expands while the contribution of the Reynolds shear stress drops, which indicates a relaminarization of the flow. However, the contribution of the particle stress also increases with $\overline{\alpha}_p$ to a point where it exceeds the Reynolds stress at $\overline{\alpha}_p=12\%$.

Contrasting the flow statistics in particle-laden channels to those in particle-free channels with matching apparent kinematic viscosity $\nu^\star$ provides a more meaningful comparison. Dispersing particles at dense concentrations has the effect of reducing the volume available to the fluid and increasing the effective fluid viscosity. Consequently, the carrier fluid can be considered to have reduced apparent density $\rho_f^\star=(1-\overline{\alpha}_p)\rho_f$ and increased apparent dynamic viscosity $\mu_f^\star=(1-\overline{\alpha}_p)^{-2.8}\mu_f$, as a first approximation. The latter relation stems from the empirical model of \citet{gibilaroApparentViscosityFluidized2007}, which we use in the present numerical simulations. For this equivalent single-phase fluid with kinematic viscosity adjusted to match the apparent kinematic viscosity $\nu^\star=\mu_f^\star/\rho_f^\star$ in the particle-laden flows, the effective friction Reynolds number  is $\Rey_{\tau}^\star = (1-\overline{\alpha}_p)^{-3.8}\Rey_{\tau}$. Thus, increasing the solid volume fraction reduces the effective friction Reynolds number $\Rey_{\tau}^\star$ and leads to lower turbulence. 

For the cases with $\overline{\alpha}_p \leq 3\%$, the increase of the suspension's apparent kinematic viscosity accounts for most of the flow modifications as evidenced by the fact that the flow statistics in the particle-laden channels agree well with those in the particle-free channels with matching $\Rey_{\tau}^\star$. In particular, the change to the bulk fluid velocity matches very well as discrepancies are less than 2\% between the particle-laden channels and the particle-free channels at the corresponding  $\Rey_{\tau}^\star$. Additionally, the stress balance agrees well, provided that the Reynolds stress from the effective particle-free fluids is compared to the sum of the Reynolds stress and particle stress from the particle-laden channels.

The cases at  $\overline{\alpha}_p = 6\%$ and 12\% display greater flow modulation than can be simply accounted for by the the increase of the suspension's apparent kinematic viscosity. In these cases, turbulence modulation through the particle feedback force plays a significant role. 
The growing discrepancy with increasing $\overline{\alpha}_p$ between the sum of the Reynolds stress and particle stress from the particle-laden channels and the Reynolds stress from the equivalent particle-free channels with matching $\Rey_{\tau}^\star$ provides evidence for the greater modulating effect of the particle feedback force. Consequently, the fluid bulk velocity is significantly lower in the particle-laden channels compared to the equivalent particle-free channels. For the case at  $\overline{\alpha}_p = 12\%$, the relative difference in bulk velocity is -16.1\%, which corresponds to a relative difference in friction coefficient of -42.2\%.
\textcolor{revision}{These conclusions are drawn from runs considering only one particle to fluid density ratio. Therefore the limit of volume fraction 3\% as accounting for modulation solely through effective kinematic viscosity may not hold when considering other particle density ratios. Additional studies can elucidate the role of density ratio in determining the limit of volume fraction induced effective viscosity dominating the flow modulation.}

\textcolor{revision}{Finally, it should be noted that some uncertainty remains regarding the exact functional dependence of the effective friction Reynolds number on the solid volume fraction. This is because the exact form of $\Rey_{\tau}^\star$ depends on the closure of the residual viscous stress tensor in Euler-Lagrange simulations, which is typically carried out with an effective vicosity model. While we have used an empirical model \citep{gibilaroApparentViscosityFluidized2007} derived from fluidized bed experiments, further work on closures of the residual viscous stress tensor is needed to dispel the uncertainty. With that regards, particle-resolved direct numerical simulations shall be a useful approach to elucidate the correct closures at high volume fractions. That said, the framework presented herein shows how effective viscosity alters the Reynolds number and thereby modulates turbulence.}

\section*{ACKNOWLEDGMENTS}
This work was supported by the donors of ACS Petroleum Research Fund under Doctoral New Investigator   Grant 62195-DNI9. M.H.K served as Principal Investigator on ACS PRF 62195-DNI9 that provided support for J.S.V.D.

\appendix
\section{Analysis of the stress balance}\label{sec:appendix_a}
Introducing particles to the channel flow leads to an additional stress in the stress balance. We perform a similar analysis as in the single phase case. Beginning by Reynolds-averaging the streamwise direction of equation (\ref{eq:momentum}) gives
\begin{equation}
\frac{d}{dy}(\langle \mu_{f}^\star \frac{\partial}{\partial y} u_{f} \rangle-\rho_f \langle \alpha_{f} u_{f}''v_{f}'' \rangle )+\langle F_{p,x} \rangle = - \langle \alpha_{f} A \rangle
\label{eq:Re_Avg_Momentum}
\end{equation}
where $\langle F_{p,x} \rangle$ is the average streamwise particle momentum exchange. Quantities with double primes indicate fluctuations with respect to the Favre averaging, that is
\begin{equation}
	\widetilde{u}_f = \frac{\langle \alpha_p u_f \rangle}{\langle \alpha_p \rangle}
\end{equation}
The velocity can then be decomposed as
\begin{equation}
	u_f = \widetilde{u}_f + u_f''
\end{equation}
Integrating equation (\ref{eq:Re_Avg_Momentum}) gives the stress balance,
\begin{equation}
	    \langle \mu_{f}^\star  \frac{\partial}{\partial y} u_{f} \rangle-\rho_f \langle \alpha_{f} u_{f}''v_{f}'' \rangle + \langle \Tau_{p} \rangle = \tau_{w} \left( 1-\frac{1}{h \overline{\alpha}_f} \int_{0}^{y} \langle \alpha_{f} \rangle dy'\right) \label{eq:stress_balance_non_linear}
\end{equation}
 where, the particle stress $\langle \Tau_{p} \rangle$ is
\begin{equation}
	\langle \Tau_{p} \rangle = \int_0^{y} \langle F_{p,x} \rangle dy' - \frac{1}{2h} \int_0^{2h} F_{p,x} dy
\end{equation}
and the wall shear stress is
\begin{equation}
	\tau_{w} = \left.\langle \mu_{f}^\star \frac{\partial}{\partial y} u_{f} \rangle \right|_{y=0}
\end{equation}
which relates to pressure gradient as following
\begin{equation}
	A = \frac {\tau_{w}} {h \overline{\alpha}_{f}}
\end{equation}

The right hand side of equation (\ref{eq:stress_balance_non_linear}) is not the familiar linear function known for constant pressure gradient forcing. To recover this behavior we define a ``stress distortion" term $\psi$ as
\begin{equation}
	\psi = \frac{ \left( 1-\frac{1}{h \overline{\alpha}_f} \int_{0}^{y} \langle \alpha_{f} \rangle dy'\right) } {\left( 1-\frac{y}{h} \right)}
\end{equation}
The final form of the stress balance with the stress distortion term is
\begin{equation}
    \label{eq:Stress_Balance}
	\langle \mu_{f}^\star  \frac{\partial}{\partial y} u_{f} \rangle-\rho_f \langle \alpha_{f} u_{f}''v_{f}'' \rangle + \langle \Tau_{p} \rangle = \tau_{w} \psi \left( 1-\frac{y}{h} \right).
\end{equation}
This stress balance analysis accounts for high particle volume fractions.
 
{\color{revision}

\section{Sensitivity of fluid statistics to model parameters.} \label{sec:appendix_c}

In this appendix,  we show that (i) variations in the restitution coefficient $e$ and (ii) the inclusion of the Saffman lift force and added mass have little impact on the fluid statistics, and thereby the mechanisms described in this study

To assess the impact of varying restitution coefficient $e$, we compare the streamwise fluid velocity and rms velocity fluctuations from two simulations with $e=0.65$ and $e=0.9$. For this comparison, we neglect the Saffman lift and added mass and retain only the effects of the undisturbed flow and the drag force in equation (\ref{eq:lpt_model}). Further, we use the inertial drag correction of \citet{tennetiDragLawMonodisperse2011}. All other numerical parameters are as described in sections \S\ref{sec:Num_experiments} and \S\ref{sec:Math_Model}.

\begin{figure}
    \begin{subfigure}{0.49\linewidth}
    \includegraphics[width=\linewidth]{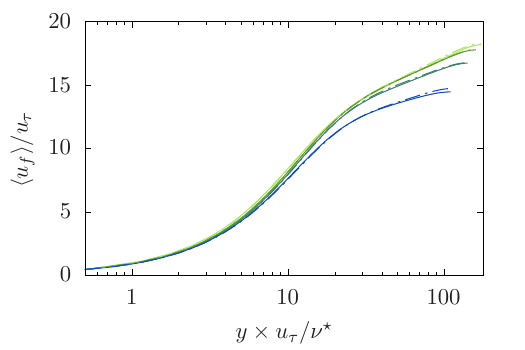}
    \caption{\label{fig:u_mean_profile_app}}
    \end{subfigure}
    \begin{subfigure}{0.49\linewidth}
    \includegraphics[width=\linewidth]{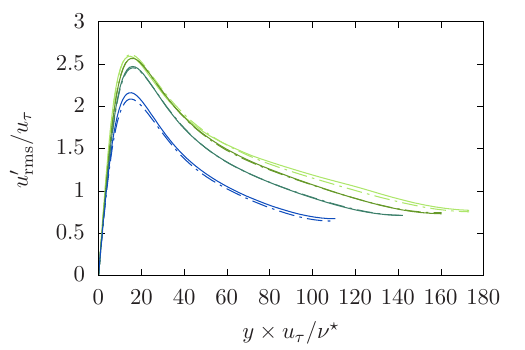}
    \caption{\label{fig:u_rms_profile_app}}
    \end{subfigure}
    \begin{subfigure}{0.49\linewidth}
    \includegraphics[width=\linewidth]{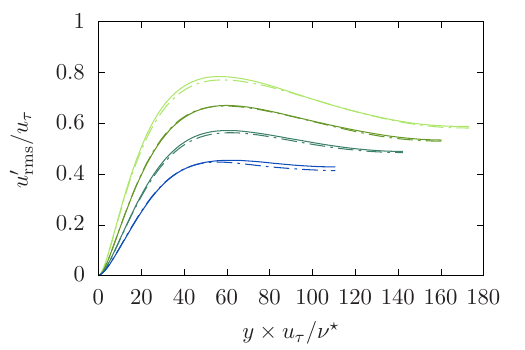}
    \caption{\label{fig:v_rms_profile_app}}
    \end{subfigure}
    \begin{subfigure}{0.49\linewidth}
    \includegraphics[width=\linewidth]{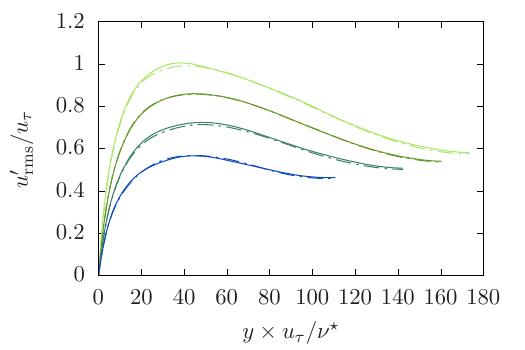}
    \caption{\label{fig:w_rms_profile_app}}
    \end{subfigure}
     \caption{\color{revision}Profiles of mean fluid velocity and rms velocity fluctuations. Here, the wall-normal position is scaled using the friction Reynolds number and the apparent kinematic viscosity. Solid lines correspond to restitution coefficient of 0.65, while dash-dotted lines correspond to restitution coefficient of 0.9 at $\overline{\alpha}_p=1\%$ (\textcolor{gpltAlpha01}{\LineStyleSolid}/\textcolor{gpltAlpha01}{\LineStyleDashDot}), $\overline{\alpha}_p=3\%$ (\textcolor{gpltAlpha03}{\LineStyleSolid}/\textcolor{gpltAlpha03}{\LineStyleDashDot}), $\overline{\alpha}_p=6\%$ (\textcolor{gpltAlpha06}{\LineStyleSolid}/\textcolor{gpltAlpha06}{\LineStyleDashDot}), and $\overline{\alpha}_p=12\%$ (\textcolor{gpltAlpha12}{\LineStyleSolid}/\textcolor{gpltAlpha12}{\LineStyleDashDot}). Fluid statistics are not sensitive to this parameter.
     \label{fig:profile_app}}
\end{figure}
Figure \ref{fig:profile_app} shows that both simulations with $e=0.65$ and $e=0.9$ yield mean streamwise fluid velocity and rms velocity fluctuation profiles that are in very close agreement. The deviations do not exceed 4 \% in any of the curves. This shows that the fluid statistics are not meaningfully sensitive to changes in the restitution coefficient.

Next, we carry a similar comparison for two simulations. In the first, we neglect the Saffman lift and added mass and use the drag correction  \citet{tennetiDragLawMonodisperse2011}. The second simulation is as described in \S\ref{sec:Math_Model}, i.e., it include the effects of Saffman lift, added mass, and the drag correction of \citet{tavanashadParticleresolvedSimulationFreely2021}. Here too, the profiles of mean streamwise fluid velocity and rms velocity fluctuations in figure \ref{fig:profile_app2} show very close agreement between the two simulations. This shows that drag and the effects of the undisturbed flow dominate the dynamics, while Saffman lift and added mass play a negligible role in the present regime. Further, the two drag corrections by \citet{tavanashadParticleresolvedSimulationFreely2021} and \citet{tennetiDragLawMonodisperse2011} do not lead to significant differences.

\begin{figure}
    \begin{subfigure}{0.49\linewidth}
    \includegraphics[width=\linewidth]{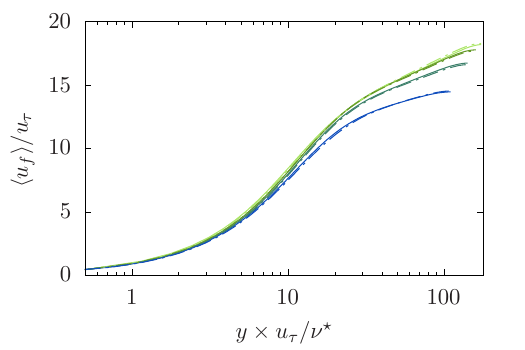}
    \caption{\label{fig:u_mean_profile_app}}
    \end{subfigure}
    \begin{subfigure}{0.49\linewidth}
    \includegraphics[width=\linewidth]{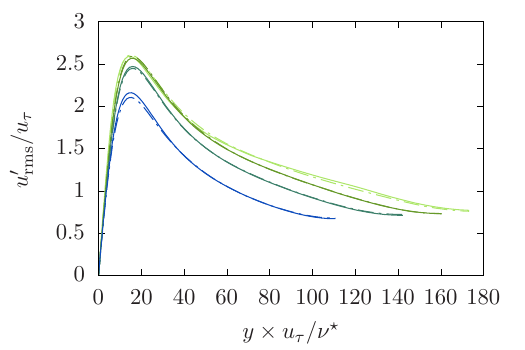}
    \caption{\label{fig:u_rms_profile_app}}
    \end{subfigure}
    \begin{subfigure}{0.49\linewidth}
    \includegraphics[width=\linewidth]{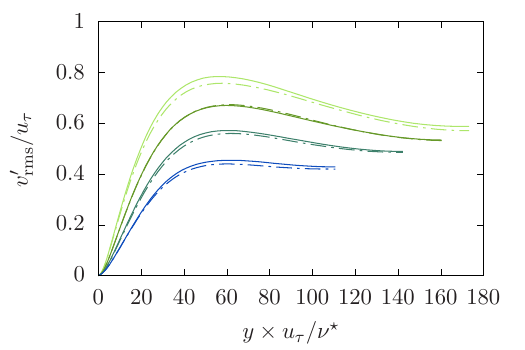}
    \caption{\label{fig:v_rms_profile_app}}
    \end{subfigure}
    \begin{subfigure}{0.49\linewidth}
    \includegraphics[width=\linewidth]{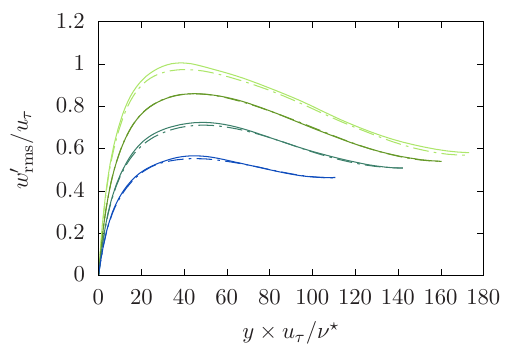}
    \caption{\label{fig:w_rms_profile_app}}
    \end{subfigure}
     \caption{\color{revision}Profiles of mean fluid velocity and rms velocity fluctuations. Here, the wall-normal position is scaled using the friction Reynolds number and the apparent kinematic viscosity. Solid lines correspond to simulations without lift, and without added mass, and using the drag correction of \citet{tennetiDragLawMonodisperse2011}. Dash-dotted lines correspond to simulations with lift and added mass, and using the drag correction of \citet{tavanashadParticleresolvedSimulationFreely2021}. $\overline{\alpha}_p=1\%$ (\textcolor{gpltAlpha01}{\LineStyleSolid}/\textcolor{gpltAlpha01}{\LineStyleDashDot}), $\overline{\alpha}_p=3\%$ (\textcolor{gpltAlpha03}{\LineStyleSolid}/\textcolor{gpltAlpha03}{\LineStyleDashDot}), $\overline{\alpha}_p=6\%$ (\textcolor{gpltAlpha06}{\LineStyleSolid}/\textcolor{gpltAlpha06}{\LineStyleDashDot}), and $\overline{\alpha}_p=12\%$ (\textcolor{gpltAlpha12}{\LineStyleSolid}/\textcolor{gpltAlpha12}{\LineStyleDashDot}). Saffman lift and added mass have a negligible effect in the present regime. Further, the two drag correction do not lead to significant differences.
     \label{fig:profile_app2}}
\end{figure}
}

{\color{revision}
\section{Grid convergence study}\label{sec:appendix_e}
To demonstrate that our results are grid independent, we present the results of a grid convergence study for case 5, which corresponds to $\overline{\alpha}_p = 0.12$. 

We compare the flow statistics from two simulations at increasing resolution. In the first one, we solve the equations of motions on a domain of size $256\times 124 \times 168$, as described in \$\ref{sec:Num_experiments}. In the second simulation, we use a higher resolution corresponding to $512\times 256 \times 336$. Thus, the number of grid points is double of the coarser simulation in each direction. 

\begin{figure}
    \begin{subfigure}{0.49\linewidth}
    \includegraphics[width=\linewidth]{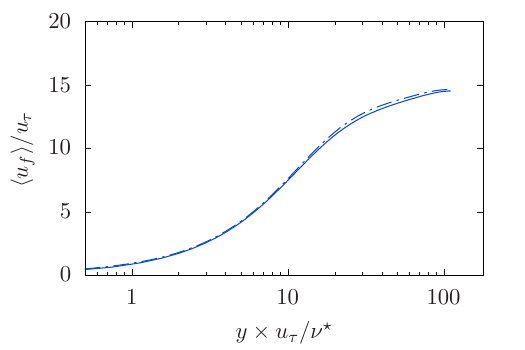}
    \caption{\label{fig:u_mean_profile_app}}
    \end{subfigure}
    \begin{subfigure}{0.49\linewidth}
    \includegraphics[width=\linewidth]{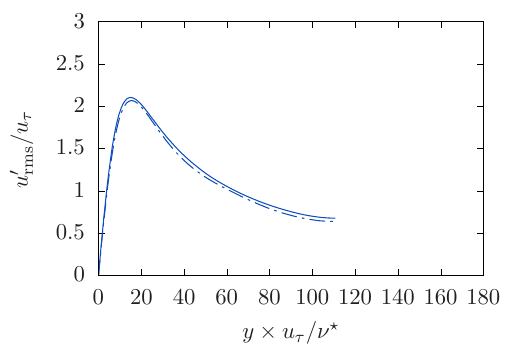}
    \caption{\label{fig:u_rms_profile_app}}
    \end{subfigure}
    \begin{subfigure}{0.49\linewidth}
    \includegraphics[width=\linewidth]{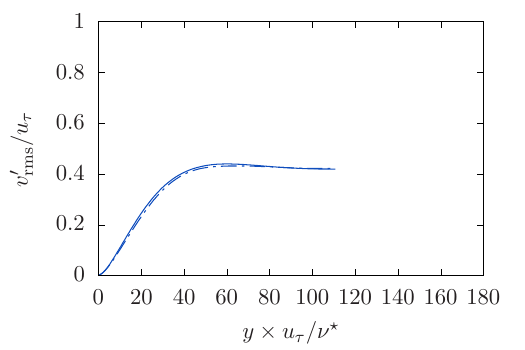}
    \caption{\label{fig:v_rms_profile_app}}
    \end{subfigure}
    \begin{subfigure}{0.49\linewidth}
    \includegraphics[width=\linewidth]{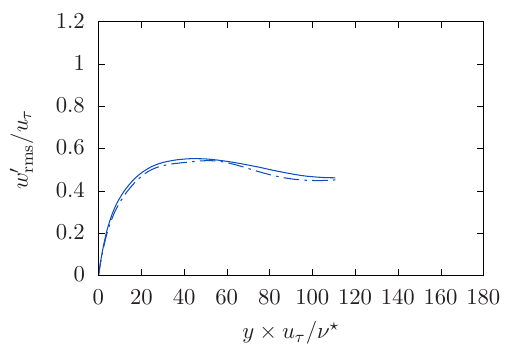}
    \caption{\label{fig:w_rms_profile_app}}
    \end{subfigure}
     \caption{\color{revision} Profiles of mean fluid velocity and rms velocity fluctuations, at volume fraction $\overline{\alpha}_p = 0.12$. Solid lines correspond to the grid described in the main document, while dashed lines correspond to the refined grid. The agreement is close, showing that the simulation is grid converged.
     \label{fig:profile_app3}}
\end{figure}

Figure \ref{fig:profile_app3} shows that increasing the resolution does not lead to significant change in the fluid statistics. Given the large number of simulations required for this study, we use the resolution  $256\times 124 \times 168$ in all remaining simulations, as it provides accurate statistics at a trackable computational cost.
}

\section{Relationship between bulk fluid velocity and stresses in a slurry}\label{sec:appendix_b}
An expression that relates the bulk fluid velocity $U_{f,b}$ to the viscous, Reynolds, and particle stresses can be derived by integrating the stress balance in equation (\ref{eq:Stress_Balance}):
\begin{equation}
	U_{f,b} = \frac{1}{h}\int_{0}^h\left\{\frac{1}{\mu^{\star}_{f}}\int_{0}^{y} \left ( \tau_{w} \psi\left ( 1-\frac{y'}{h}\right ) + \rho_{f} \langle \alpha_{f} u''_f v''_f \rangle - \langle \Tau_p \rangle \right )dy'\right\}dy. \label{eq:B1}\end{equation}

To simplify the expression above, we consider that variations of fluid volume fraction $\alpha_f$ and the distortion term $\psi$ are negligible. With regards to $\alpha_f$, limited particle clustering observed in figure \ref{fig:vfp_profile_a} and relatively low particle volume fraction lead to variations of fluid volume fraction $\alpha_f=1-\alpha_p$ that do not exceed 2.5\% in any case. Thus, we consider that $\alpha_f\simeq \overline{\alpha}_f$ in the following analysis. Further, as is shown in figure \ref{fig:Psi_variation}, $\psi$ is nearly constant and equal to 1 throughout the channel.

\begin{figure}
  \centering
    \includegraphics[width=5in,clip]{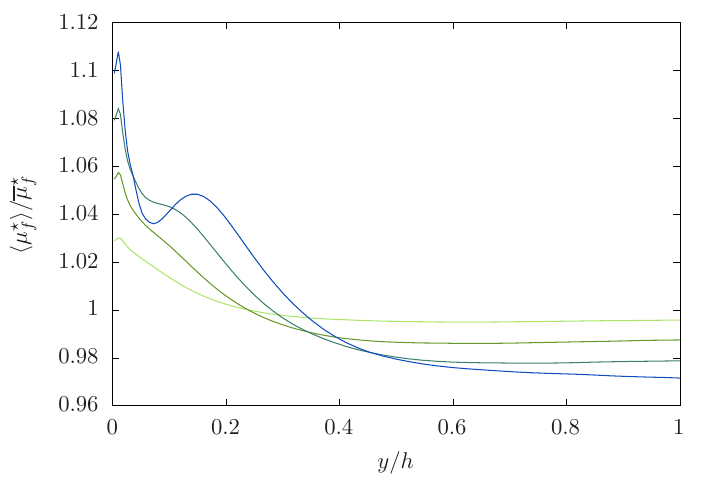}
  \caption{Variation of the normalized apparent viscosity with wall-normal distance at volume fraction $\overline{\alpha}_p=1\%$ (\textcolor{gpltAlpha01}{\LineStyleSolid}), $\overline{\alpha}_p=3\%$ (\textcolor{gpltAlpha03}{\LineStyleSolid}), $\overline{\alpha}_p=6\%$ (\textcolor{gpltAlpha06}{\LineStyleSolid}), and $\overline{\alpha}_p=12\%$ (\textcolor{gpltAlpha12}{\LineStyleSolid}).
  \label{fig:Mu_variation}}
\end{figure}

Figure \ref{fig:Mu_variation} shows that variations of $\mu_f^\star$ can also be neglected. While increasing the particle concentration leads to significant increase in the overall apparent viscosity, particle clustering does not lead to significant fluctuations of the local effective viscosity $\mu_f^{\star}$. The apparent viscosity normalized the channel-averaged value $\overline{\mu}_f^{\star}$ shows a peak in the near wall region resulting from a corresponding peak in particle volume fraction. The local effective viscosity peaks represent a departure from the channel-averaged values of \textcolor{revision}{3.0\%, 5.7\%, 8.4\%, and 10.0\%} at particle volume fractions 1\%, 3\%, 6\%, and 12\%, respectively.  Towards the channel center the effective viscosity plateaus to a value slightly lower than the channel-averaged value. In all cases, the relative deviations of $\mu_f^{\star}/\overline{\mu}_f^{\star}$ towards the center do not exceed 3\%. 

With the assumptions above, equation (\ref{eq:B1}) can be simplified and written as
\begin{equation}
	\frac{U_{f,b}}{u_{\tau}} \simeq \frac{\Rey^{\star}_{\tau}}{3}\left ( 1  + \frac{3}{(u_{\tau} h)^2}\int^h_0 \int_0^y \langle u''_f v''_f \rangle - \frac{M}{ \rho_p \overline{\alpha}_p}\langle \Tau_p \rangle dy'dy \right).
\end{equation}

%\bibliography{references/jon,references/houssem}
\bibliography{references/houssem,references/jon}
\end{document}